\documentclass[journal=jcisd8,manuscript=article,titles=true]{achemso}

\usepackage{chemformula} 
\usepackage[T1]{fontenc} 

\usepackage{newfloat}
\DeclareFloatingEnvironment[name={Supplementary Table},fileext=lst,listname={List of Supplementary Tables}]{supptable}
\DeclareFloatingEnvironment[name={Supplementary Figure},fileext=lof]{suppfigure}

\usepackage{color}

\usepackage{amsmath,amsfonts,amsthm}
\usepackage{chngcntr}
\usepackage{gensymb}
\usepackage{amssymb}

\usepackage{soul}
\usepackage[T1]{fontenc} 
\usepackage[english]{babel} 
\usepackage{multirow}
\usepackage{graphicx} 
\usepackage{lmodern}
\usepackage{mathtools}
\usepackage{hyperref}   

\usepackage{lipsum} 
\usepackage{fancyhdr} 
\numberwithin{equation}{section} 
\numberwithin{figure}{section} 
\numberwithin{table}{section} 

\counterwithout{figure}{section}
\counterwithout{table}{section}

\title{Harnessing Linear and Nonlinear Optical Responses in Ferroelectric LaMoN$_3$ for Enhanced Photovoltaic Efficiency}

\author{Surajit Adhikari}
\affiliation{Department of Physics, Indian Institute of Technology Bombay, Mumbai 400076, India.}
\altaffiliation{S.A and S.S.P contributed equally to this work.}

\author{Sanika S. Padelkar}
\affiliation{Department of Physics, Indian Institute of Technology Bombay, Mumbai 400076, India.}
\alsoaffiliation{School of Chemistry, Monash University, Victoria 3800, Australia.}
\alsoaffiliation{Department of Materials Science \& Engineering, Monash University, Victoria 3800, Australia.}
\alsoaffiliation{IITB-Monash Research Academy, Indian Institute of Technology Bombay, Mumbai 400076, India.}
\altaffiliation{S.A and S.S.P contributed equally to this work.}

\author{Jacek J. Jasieniak}
\affiliation{Department of Materials Science \& Engineering, Monash University, Victoria 3800, Australia.}

\author{Alexandr N. Simonov}
\affiliation{School of Chemistry, Monash University, Victoria 3800, Australia.}

\author{Aftab Alam}
\affiliation{Department of Physics, Indian Institute of Technology Bombay, Mumbai 400076, India.}
\alsoaffiliation{IITB-Monash Research Academy, Indian Institute of Technology Bombay, Mumbai 400076, India.}
\email{aftab@iitb.ac.in}


\begin{document}

\begin{tocentry}
\begin{center}
\includegraphics[height=4.45cm,width=8.25cm]{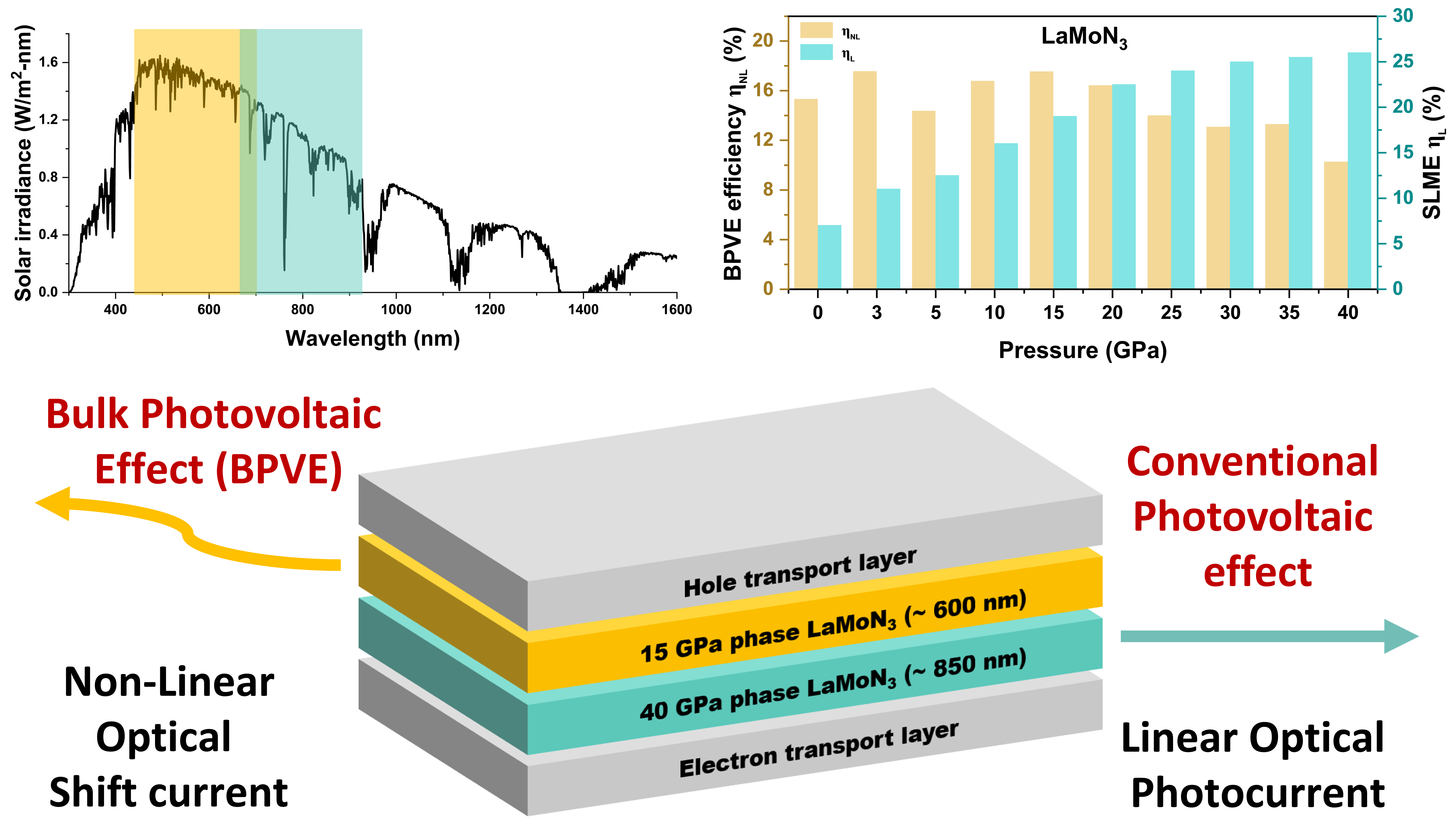}
\end{center}
\end{tocentry}

\begin{abstract}
  Nitride perovskites are an emerging class of materials predicted to exhibit diverse functional properties, yet remain underexplored due to synthesis challenges of oxygen-free nitrides. Recently, LaMoN$_{3}$ has been reported as an oxygen-free nitride perovskite with polar symmetry, exhibiting excellent dynamic stability and ferroelectric properties under moderate pressure. However, its phase stability, linear and non-linear optical response, excitonic and polaronic behavior, and efficiency under high pressure remain unexplored. Applying pressure enables systematic tuning of the electronic structure properties, thereby facilitating the identification of phases optimized for either linear or nonlinear optical responses. Therefore, in this work, we systematically investigate these properties of LaMoN$_{3}$ up to 40 GPa using first-principles methods, including density functional theory, density functional perturbation theory, many-body perturbation theory (namely G$_{0}$W$_{0}$ and BSE), and tight binding approximation (TBA) model. Our study shows that LaMoN$_{3}$ remains dynamically stable and retains its single-phase structure up to 40 GPa. The compound exhibits an indirect bandgap that decreases from 2.17 eV (0 GPa) to 1.45 eV (40 GPa) at the G$_{0}$W$_{0}$@PBE level. Using the BSE, we find that pressure enhances the SLME while lowering the exciton binding energy, both favorable for photovoltaic applications. Additionally, carrier-phonon interactions, analyzed via the Fr\"ohlich model, are found to strengthen under pressure, resulting in reduced carrier mobility. Complementing these findings, bulk photovoltaic (BPV) and non-linear effects were studied. The BPV efficiency trend with pressure mimics the behavior of the shift current density \(J_{SC}\), peaking near 15 GPa before declining at higher pressures due to a diminished nonlinear shift current response. These results highlight pressure-tuned regimes to enhance photovoltaic performance. We thereby propose multi-junction device, combining absorber layers optimized for linear and nonlinear optical currents: the 40 GPa phase enables efficient linear response in micrometer-scale absorbers, while the 15 GPa phase maximizes nonlinear current in nanometer-scale absorbers, together boosting solar energy conversion through complementary mechanisms.

\end{abstract}

\maketitle

\section{Introduction} 
Perovskites, with the general formula ABX$_{3}$, represent a versatile
class of materials characterized by a three-dimensional (3D) network
of corner-sharing BX$_{6}$ octahedra, within which the A-site cation
resides in the cavities formed by the octahedra\citep{Ref.1}. These
materials display a remarkable range of functional properties, including
ferroelectricity, ferromagnetism, piezoelectricity, and exceptional
optoelectronic behavior, making them highly attractive for fundamental
research and advanced technological applications\citep{Ref.2,Ref.3}.
Oxide perovskites (X = O$^{2-}$), like Pb(Zr, Ti)O$_{3}$ and BaTiO$_{3}$,
have been extensively utilized in electrochemical cells, microelectromechanical
actuators, and ceramic capacitors owing to their strong piezoelectric
responses, but their large bandgaps limit optoelectronic performance
\citep{Ref.4,Ref.5,Ref.6}. In contrast, halide perovskites APbX$_{3}$
(A = Cs$^{+}$, CH$_{3}$NH$_{3}^{+}$; X = Cl$^{-}$, Br$^{-}$,
I$^{-}$) have achieved remarkable success in solar cells, LEDs, and
photodetectors, reaching power conversion efficiencies above 27\%
\citep{Ref.7,Ref.8}. However, their practical deployment faces
challenges due to the toxicity of lead (Pb) and the instability arising
from their volatile and hygroscopic lattice. Recently, lead-free
chalcogenide perovskites ABX$_{3}$ (A = Ca$^{2+}$, Sr$^{2+}$, Ba$^{2+}$;
B = Zr$^{4+}$, Hf$^{4+}$, Sn$^{4+}$; X = S$^{2-}$, Se$^{2-}$)
have been proposed as alternatives with excellent optoelectronic
properties \citep{Ref.9,Ref.10}, but face challenges such as limited
synthesis reports, defect-induced carrier losses, and relatively low
device efficiencies. These drawbacks motivate the search for other
lead-free, stable perovskite-based or perovskite-inspired materials
with superior optoelectronic performance.

Recently, nitride compounds have garnered increasing attention due
to their mixed covalent-ionic bonding nature, arising from the
moderate electronegativity of nitrogen ($\chi_{N}$ = 3.0). This unique
bonding character imparts superior solar absorption and electrical
transport properties compared to the extensively studied oxides\citep{Ref.49}. Moreover,
nitrides present significant interest from a fundamental science perspective,
as they represent a relatively unexplored yet highly diverse class
of inorganic compounds \citep{Ref.11}. Several theoretical studies
have explored substituting the X-site anion from group VI or VII elements
with group V elements (specifically nitrogen) to assess the feasibility
of forming nitride perovskites. For example, ABN$_{3}$ (A = La, Ce,
Eu, Yb; B = W, Re) compounds have been predicted to be thermodynamically
stable \citep{Ref.12}. A few nitride perovskites, such as LaWN$_{3}$
and ThTaN$_{3}$, have been successfully synthesized experimentally
\citep{Ref.13,Ref.14}. Gui et al. theoretically investigated LaMoN$_{3}$,
revealing its superior ferroelectric properties \citep{Ref.15}. Furthermore,
CeWN$_{3}$ and CeMoN$_{3}$ were identified to host promising features through high-throughput
computational screening and subsequently fabricated using thin-film
growth techniques \citep{Ref.16}.

\textcolor{black}{Building on these foundational advances, nitride perovskites have emerged as promising candidates for photovoltaic (PV) applications, primarily due to their enhanced chemical robustness stemming from strong metal-nitrogen bonds. Theoretical/computational studies have systematically evaluated their optoelectronic properties, revealing direct or quasidirect bandgaps in the optimal PV range of 1.1--2.5 eV for compounds like LaWN$_3$ ($E_g \approx 1.8$ eV), CeTaN$_3$ ($E_g \approx 1.1$--2.0 eV), and Pna2$_1$-LaWN$_3$ ($E_g \approx 1.3$ eV), alongside moderate absorption coefficients and low exciton binding energies that favor efficient carrier generation \citep{Sun2023,Ha2022,Lai2024}. High-throughput density functional theory (DFT) screenings have expanded the chemical design space, identifying over 100 thermodynamically accessible ABN$_3$ candidates with favorable effective masses and defect tolerances, positioning them as wide-bandgap top cells in perovskite/silicon or all-nitride tandems \citep{Sun2023,Ghosh2024,Lai2024}. Experimentally, oxygen-free LaWN$_3$ thin films have been synthesized via reactive sputtering, exhibiting polar symmetry, effective piezoelectric coefficients ($d_{33} \sim 40$ pm/V), and compatibility with GaN-based epitaxy for hybrid PV architectures \citep{Hines2021}. Ab-initio simulation of nitride-perovskite tandems demonstrate operational stabilities exceeding 10,000 hours under AM1.5 illumination, leveraging nitride robustness against cosmic radiation for space applications \citep{Kisi2023}. Despite these strides, challenges persist in scalable synthesis - requiring high-pressure ammonolysis or plasma-assisted methods and experimental PV devices remain in the proof-of-concept stage, with ongoing machine learning-guided efforts accelerating discovery of defect-mitigated compositions \citep{Ghosh2024,Sun2023b}. Thus, nitride perovskites complement halide systems by enabling durable, lead-free PV technologies with tailored bandgaps for multi-junction cells.}

Among the nitride perovskites, LaMoN$_{3}$ (space group R3c) is predicted
to have a moderate band gap of 1.96 eV (HSE06) and to exhibit ferroelectricity
with a spontaneous polarization of approximately 80.3 $\mu$C/cm$^{2}$
\citep{Ref.15}, exceeding that of the same phase of LaWN$_{3}$ (66
$\mu$C/cm$^{2}$ \citep{Ref.17}). The incorporation of a 4$d$ Mo
cation, whose $d$ orbitals are more localized than the spatially
extended 5$d$ orbitals of W in LaWN$_{3}$, leads to a wider band
gap, thereby enhancing the prospects for observing ferroelectricity.
Gui et al. \citep{Ref.15} also demonstrated that the nonpolar, nonperovskite
C2/c phase of LaMoN$_{3}$ undergoes a pressure-induced phase transformation
(at 1.3 GPa) into the ferroelectric perovskite R3c phase and this
phase retains pronounced ferroelectric features up to 40 GPa.

However, the phase stability and bulk photovoltaic response (particularly
the shift current arising from the nonlinear optical effects) of the non-centrosymmetric ferroelectric LaMoN$_{3}$
(R3c phase) under elevated pressures remain largely unexplored. The
shift current mechanism in the ferroelectric bulk photovoltaic effect
(BPVE) not only connects fundamental aspects of polarization, symmetry,
and nonlinear optics but also offers practical potential for high-voltage,
bias-free, and tunable BPVE-based optoelectronic devices, especially
in materials where polarization and band structure can be engineered
together. Gui et al. \citep{Ref.15} also reported that the HSE06
band gap of LaMoN$_{3}$ (R3c phase) decreases to 1.40 eV at 40 GPa,
suggesting that this compound could be promising for conventional
optoelectronic applications under high-pressure conditions. Despite
this promising electronic tunability, a comprehensive understanding
of the high-pressure behavior of LaMoN$_{3}$ requires further investigation of
its optical response, excitonic effects, and polaronic dynamics, which have not yet been done. Optical
properties, such as dielectric function and absorption coefficient,
determine the efficiency of light harvesting and emission. Excitonic
behavior, particularly the exciton binding energy and its variation
with pressure, plays a pivotal role in controlling charge carrier
separation and recombination rates, directly affecting device performance.
Polaronic effects, which stem from the interaction between charge
carriers and lattice vibrations, can significantly influence charge
transport by altering carrier mobility, recombination lifetime, and
thermal stability. Since these factors collectively govern the overall
performance of any optoelectronic material, a systematic
study of LaMoN$_{3}$ under high-pressure conditions is essential
to fully assess its potential for practical applications.

In this work, we have systematically investigated the pressure-induced
phase stability, electronic structure, optical response, excitonic
dynamics, polaronic effects, and nonlinear shift current behavior of the non-centrosymmetric
ferroelectric LaMoN$_{3}$ (R3c phase) perovskite within the framework
of density functional theory (DFT) \citep{Ref.18,Ref.19}, density
functional perturbation theory (DFPT) \citep{Ref.20}, and many-body
perturbation theory (MBPT) \citep{Ref.21,Ref.22}. Our work provides the first complete GW-BSE, polaron, and nonlinear shift-current analysis of any nitride perovskite under structural tuning. Our study reveals
that the LaMoN$_{3}$ compound exhibits excellent dynamical stability
and retains its single phase up to 40 GPa. Our findings reveal that LaMoN$_{3}$ exhibits an indirect bandgap that decreases with increasing pressure, with the G$_{0}$W$_{0}$@PBE value reducing from 2.17 eV at 0 GPa to 1.45 eV at 40 GPa. Leveraging the Bethe-Salpeter equation (BSE) approach, we find that increasing pressure enhances the spectroscopic limited maximum efficiency (SLME) \citep{Ref.39} while reducing the exciton binding energy, both of which are favorable for photovoltaic applications.  Furthermore, the carrier-phonon coupling strength and polaron mobility are calculated using the Fr\"ohlich model\citep{Ref.9,Ref.36} and Hellwarth polaron model\citep{Ref.44}, respectively. Our study suggests that carrier-phonon interactions strengthen with increasing pressure, leading to a reduction in carrier mobility. This behavior indicates that the polar nature of LaMoN$_{3}$ is enhanced under higher pressure conditions. Since LaMoN$_{3}$ crystallizes in a non-centrosymmetric phase, it exhibits intrinsic polar properties that are fundamental to its ferroelectric and photovoltaic behaviors. This non-centrosymmetric structure enables a spontaneous polarization that is pivotal for the shift current generation and related nonlinear optical effects. Our investigation reveals that the shift current photovoltaic efficiency of LaMoN$_{3}$ increases slightly with pressure up to 15 GPa but decreases at higher pressures, mirroring the behavior of the short-circuit current density. The decline at elevated pressure is attributed to reduced material polarization, which diminishes the nonlinear optical shift current response, thereby reducing overall device efficiency. The synergy between linear (SLME) and nonlinear (shift-current) photovoltaic mechanisms identified here, represents a novel concept, uniquely enabled by our calculations.

\section{Computational details}

First-principles DFT,
DFPT, and MBPT calculations were performed using the Vienna ab initio
simulation package (VASP) \citep{Ref.23,Ref.24}. The projector augmented-wave
(PAW) method was employed to accurately describe the interactions
between valence electrons and the atomic core \citep{Ref.25}. The
PAW pseudopotentials included the following valence electron configurations
for each element: 6s$^{2}$5d$^{1}$ for La, 5s$^{1}$4d$^{5}$ for
Mo, and 2s$^{2}$2p$^{3}$ for N, respectively. The exchange-correlation
(xc) functional of Perdew, Burke, and Ernzerhof (PBE), based on the
generalized gradient approximation (GGA), was used to treat electron-electron
interactions \citep{Ref.26}. Structural optimizations were carried
out using the conjugate-gradient method,
with an optimized plane-wave cutoff energy of 520 eV. The convergence
thresholds for self-consistent field iterations and geometry relaxations
were set to 10$^{-6}$ eV for total energy and 0.01 eV/\textup{\AA}
for the Hellmann-Feynman forces. A $\Gamma$-centered $8\times8\times8$
k-point mesh with the PBE functional was used for structural optimization.
It is important to note that the studied system LaMoN$_3$ is essentially a non-magnetic system, as none of the magnetic calculations (ferromagnetic, antiferromagnetic and a few ferrimagnetic) converge to the desired magnetically ordered phase but rather relxes to the non-magnetic ground state, as also reported elsewhere \citep{Ref.15}.  This is true even under applied
external pressure\citep{Ref.15}. The optimized crystal structures
were visualized using the VESTA (Visualization for Electronic and
STructural Analysis) software \citep{Ref.27}. Phonon dispersion curves
were obtained using the DFPT method employing $2\times2\times2$ supercells,
as implemented in the PHONOPY package \citep{Ref.28}, while the acoustic
sum rule (ASR) was enforced using the HIPHIVE package \citep{Ref.29}. We have included the mathematical formulation of the ASR in the Supporting Information (SI).

Since the PBE functional typically underestimates band gaps due to
self-interaction errors, more accurate electronic structures were
simulated using the hybrid Heyd-Scuseria-Ernzerhof (HSE06) functional
\citep{Ref.30} and MBPT-based G$_{0}$W$_{0}$@PBE \citep{Ref.33,Ref.34}
calculations. Optical properties were evaluated by solving
the MBPT-based BSE \citep{Ref.31,Ref.32}
on top of G$_{0}$W$_{0}$@PBE calculations. \textcolor{black}{GW-BSE calculations were carried out using a $\Gamma$-centered $4\times4\times4$ $\mathrm{k}$-point mesh, with the quasiparticle energies converged using 800 unoccupied states. In the BSE calculations, 8 valence and 8 conduction bands were included, which were sufficient to ensure convergence of the electron–hole interaction kernel and the resulting excitonic properties.} Additionally,
the ionic contribution to the dielectric function was obtained via
DFPT\citep{Ref.20} with
a $8\times8\times8$ $\mathrm{k}$-point grid.

\begin{figure*}[t]  
\centering
\includegraphics[width=1.0\linewidth]{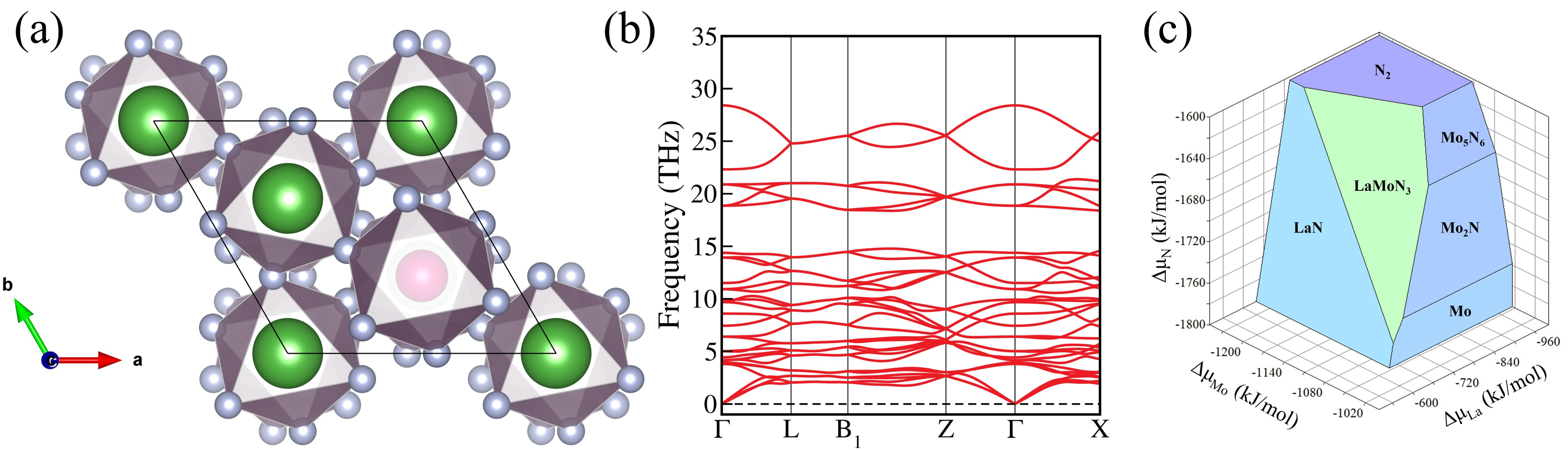}
\centering
\caption{\label{fig:1}(a) Crystal structure, (b) phonon dispersion at 0 GPa, and (c) chemical phase diagram  at 0 GPa for LaMoN$_{3}$. In panel (a), the green, pink, and silver spheres represent La, Mo, and N atoms, respectively.} 
\end{figure*}

The chemical phase diagrams of LaMoN$_{3}$ at ambient and under all the external pressures were obtained using the Chesta code\cite{key_1} by calculating the Gibbs free energies\cite{therrien2023vaspgibbs} of the target systems and its corresponding secondary phases using DFPT\citep{Ref.20}. The electron relaxation time was determined using the AMSET software\cite{AMSET}, by incorporating two scattering mechanisms: acoustic phonon scattering (ADP) and polar optical phonon scattering (POP) which are dominant. The other scattering contributions (piezoelectric and ionized impurity) are negligibly small. This tool employs the momentum relaxation-time approximation within the Boltzmann transport equation framework to compute scattering rates and carrier mobilities.  Isosurface plots of band decomposed partial charge density at highest occupied molecular orbital (HOMO) and lowest unoccupied molecular orbital (LUMO) were obtained using wavefuctions converged with a $\Gamma$-centered $14\times14\times14$ $\mathrm{k}$-point grid. Spin-orbit coupling (SOC) effects are small and hence are not included in the present sets of calculations. For example, in our study, the PBE band gap of LaMoN$_{3}$ at 0 GPa is 1.19 eV, which decreases only slightly to 1.14 eV upon inclusion of SOC \textcolor{black}{(see Fig. S4 of SI)}.

For nonlinear optical (NLO) properties, we employ the Wannier interpolated model\cite{ibanez2018ab}, where pyw90 and WANNIER90 \cite{mostofi2014updated} codes are used to obtain the Wannier functions. To compute the shift current response, maximally localized Wannier functions were generated using Wannier90 from the VASP-calculated band structure. The Wannier interpolation method enabled the use of dense \(\mathbf{k}\)-point meshes, up to \(100 \times 100 \times 100\), which are necessary for achieving convergence in the nonlinear optical response calculations. The frequency-dependent shift current conductivity tensor \(\sigma_{ijk}(\omega)\) was evaluated within the length gauge formalism, incorporating Berry connection matrix elements and their derivatives obtained from the Wannier Hamiltonian. The energy-conserving \(\delta\)-function was approximated by a Gaussian broadening with a width of 0.04 eV.  Finally, the calculated shift current conductivity tensors were integrated over the solar spectrum to obtain the shift current density J$_{SC}$ and power conversion efficiencies were evaluated by adapting the shift current photovoltaic efficiency model proposed by Sauer et al..\cite{sauer2023shift}

\section{Results and Discussion}

\subsection{Structural Properties}

LaMoN$_{3}$ crystallizes in
the rhombohedral R3c space group (No. 161), which is a non-centrosymmetric
polar structure. This
space group belongs to the trigonal crystal system, with the corresponding
crystal structure shown in  Fig. \ref{fig:1}(a).
The unit cell of LaMoN$_{3}$ contains two La atoms, two Mo atoms,
and six N atoms, where the Mo and N atoms form a corner-sharing distorted
octahedral network of MoN$_{6}$ with an average Mo$-$N bond length
of 2.07 \textup{~\AA} under ambient conditions. \textcolor{black}{The term “perovskite” is frequently used in a broader sense to include distorted and derived structures beyond the ideal corner-sharing octahedral network. In this context, R3c phases exhibiting partial face-sharing connectivity, such as LaMoN$_3$, are commonly classified as distorted perovskites when they originate from the parent framework under structural distortions or external perturbations. Accordingly, we describe R3c LaMoN$_3$ as a distorted perovskite, consistent with prior literature.\citep{Ref.15}}  Gui et al. \citep{Ref.15}
also reported that the R3c phase becomes more stable at higher pressures
($\sim$ 1.5 GPa) than at ambient conditions. Therefore, in our study,
we applied hydrostatic pressures up to 40 GPa and investigated the
phase stability under varying pressures, as discussed later. Table \ref{tab:1} shows the calculated
lattice parameters corresponding to different pressures applied,
indicating that the a and b axes exhibit greater resistance to compression
than the c axis.

\begin{table*}
\caption{\label{tab:1}Calculated lattice parameters (a, b, and c) and octahedral
distortion parameters of MoN$_{6}$ octahedra\textemdash including
the average bond length ($\mathrm{\mathit{d}_{0}}$), bond angle variance
($\sigma^{2}$), polyhedral volume ($\mathrm{V_{o}}$), and bond length
distortion index ($\mathrm{D}$) under different applied pressures for LaMoN$_3$.}

\begin{centering}
\begin{tabular}{cccccccc}
\hline 
\multirow{2}{*}{Pressure (GPa)} & \multicolumn{2}{c}{Lattice parameters} &  & \multicolumn{4}{c}{Octahedral distortion parameters}\tabularnewline
\cline{2-3}\cline{5-8}
 & a = b (\textup{\AA}) & c (\textup{\AA}) &  & $\mathrm{\mathit{d}_{0}}$ (\textup{\AA}) & $\sigma^{2}$ (deg$^{2}$) & $\mathrm{V_{o}}$ (\textup{\AA}$^{3}$) & $\mathrm{D}$\tabularnewline
\hline 
0 & 5.71 & 14.05 &  & 2.07 & 70.23 & 11.44 & 0.0986\tabularnewline
3 & 5.68 & 13.90 &  & 2.05 & 64.79 & 11.20 & 0.0913\tabularnewline
5 & 5.66 & 13.82 &  & 2.04 & 61.10 & 11.06 & 0.0868\tabularnewline
10 & 5.62 & 13.64 &  & 2.02 & 54.63 & 10.76 & 0.0777\tabularnewline
15 & 5.59 & 13.50 &  & 2.00 & 50.61 & 10.51 & 0.0706\tabularnewline
20 & 5.56 & 13.38 &  & 1.99 & 46.90 & 10.30 & 0.0651\tabularnewline
25 & 5.53 & 13.28 &  & 1.98 & 44.47 & 10.12 & 0.0601\tabularnewline
30 & 5.50 & 13.18 &  & 1.97 & 42.06 & 9.95 & 0.0561\tabularnewline
35 & 5.48 & 13.09 &  & 1.96 & 40.90 & 9.80 & 0.0527\tabularnewline
40 & 5.45 & 13.02 &  & 1.95 & 39.98 & 9.66 & 0.0496\tabularnewline
\hline 
\end{tabular}
\par\end{centering}
\end{table*}

To gain deeper insights into the pressure-induced variations in structural
parameters, we evaluated the octahedral distortion parameters including
the average bond length ($\mathrm{\mathit{d}_{0}}$), bond angle variance
($\sigma^{2}$), polyhedral volume ($\mathrm{V_{o}}$), and bond length
distortion index ($\mathrm{D}$) for the MoN$_{6}$ octahedra, as
summarized in Table \ref{tab:1}. These parameters are calculated
using the following expressions \citep{Ref.10,Ref.36}:

\begin{equation}
\mathrm{Average\;bond\;length\;(\mathit{d}_{0})=\frac{1}{6}}\sum_{i=1}^{6} d_i
\end{equation}

\begin{equation}
\mathrm{Bond\;angle\;variance\;(\sigma^{2})=\frac{1}{11}\sum_{i=1}^{12}(\theta_{i}-90)^{2}}
\end{equation}

\begin{equation}
\mathrm{Distortion\;index\;(D)=\frac{1}{6}}\sum_{i=1}^{6}\frac{\left|d_{i}-d_{0}\right|}{d_{0}}
\end{equation}

where $d_{i}$ represents the individual Mo$-$N bond lengths and
$\theta_{i}$ denotes the N$-$Mo$-$N bond angles within the MoN$_{6}$
octahedra. From Table \ref{tab:1}, it is evident that all these parameters
decrease with increasing external pressure. Reduction
in the bond length distortion index of the octahedra is a measure of enhanced stability, and hence the application of external pressure indeed  helps to improve the structural stability.

Next, we simulated the dynamical stability of the investigated
perovskite under different pressures, as this property is a key indicator
of its structural integrity and functional viability.  A dynamically
stable material is characterized by real, positive phonon frequencies in the phonon dispersion
across the entire Brillouin zone, whereas imaginary (negative) frequencies
signal possible structural instabilities. The phonon spectra are obtained
from self-consistent calculations using the DFPT method \citep{Ref.20}.
The computed phonon dispersion curve at zero pressure is shown in
Fig. \ref{fig:1}(b), while the corresponding curves at elevated
pressures are presented in Fig. S1. Across all examined pressures,
the absence of imaginary phonon modes confirms that LaMoN$_{3}$ maintains
dynamical stability from 0 GPa up to 40 GPa, underscoring its robustness
at all external compression. Previous theoretical studies
by Gui et al. \citep{Ref.15} and Wang et al. \citep{Ref.37} reported
that the phonon dispersion curve of LaMoN$_{3}$ shows negative frequencies
near the $\Gamma$ point at 0 GPa. In contrast, our results indicate
that, upon enforcing the acoustic sum rule, these negative frequencies
are eliminated. From a group-theoretical perspective, the structural
symmetry of LaMoN$_{3}$ results in 30 distinct phonon modes in its
primitive cell containing 10 atoms. Among these, three are acoustic
modes, while the remaining 27 are optical modes, which can be further
classified into low- and high-frequency phonons. A particularly notable
trend emerging from our analysis is the pressure dependence of the
optical phonon branches, where the highest
optical phonon frequencies systematically shift to higher values as the pressure increases.
This stiffening of the optical modes can be attributed to the reduction
in interatomic distances under compression, which strengthens the
chemical bonds and raises the vibrational energies. Such behavior
has important implications for the thermodynamic and transport properties
of the material, as higher-frequency phonons often influence the heat
capacity, thermal expansion, and scattering mechanisms relevant to
the performance of the optoelectronic device.

It is to be noted that LaWN$_3$ is an isostructural compound which belongs to the same family as LaMoN$_3$ and has already been experimentally synthesized earlier\cite{Ref.13}. We simulated the phonon dispersion of LaWN$_3$, and observed a small negative phonon frequencies at/near the $\Gamma$-point similar to LaMoN$_3$ before applying the acoustic sum rule corrections (see Fig. S2 of SI), which disappears upon enforcing this correction. This confirms that the ASR-corrected phonon spectra provide a reliable representation of dynamical stability.

\begin{table*}[t]
\caption{\label{tab:2}Bandgaps (in eV) of LaMoN$_{3}$ at different pressures
calculated using PBE, HSE06, and G$_{0}$W$_{0}$@PBE methods,
along with the effective masses of electrons ($m_{e}^{*}$), holes
($m_{h}^{*}$) and effective mass ratio ($M=m_{h}^{*}/m_{e}^{*}$). $E_{g}^{dir}$ is the direct band gap calculated using G$_{0}$W$_{0}$@PBE, as indicated in Fig. \ref{Figure-2}(d,e,f).  All values of the
effective masses are in terms of free-electron mass ($m_{0}$). }

\centering{}%
\begin{tabular}{ccccccccc}
\hline 
\multirow{1}{*}{Pressure (GPa)} &  & PBE & HSE06 & G$_{0}$W$_{0}$@PBE & $E_{g}^{dir}$ & $m_{e}^{*}$ & $m_{h}^{*}$ & $M$\tabularnewline
\hline 
0 &  & 1.19 & 1.95 & 2.17  & 2.47 & 0.738 & 1.215 & 1.65\tabularnewline
3 &  & 1.15 & 1.88 & 2.07  & 2.32  & 0.760 & 1.178 & 1.55\tabularnewline
5 &  & 1.12 & 1.83 & 2.00  & 2.25 & 0.771 & 1.000 & 1.30\tabularnewline
10 &  & 1.06 & 1.75 & 1.87  & 2.08 & 0.870 & 1.463 & 1.68\tabularnewline
15 &  & 0.99 & 1.65 & 1.78  & 1.95 & 1.052 & 1.699 & 1.61\tabularnewline
20 &  & 0.93 & 1.59 & 1.72  & 1.84 & 1.089 & 2.047 & 1.88\tabularnewline
25 &  & 0.88 & 1.52 & 1.63  & 1.74 & 1.151 & 2.065 & 1.79\tabularnewline
30 &  & 0.84 & 1.46 & 1.55  & 1.67 & 1.286 & 3.100 & 2.41\tabularnewline
35 &  & 0.81 & 1.42 & 1.49  & 1.63 & 1.412 & 4.417 & 3.13\tabularnewline
40 &  & 0.79 & 1.39 & 1.45  & 1.57 & 1.493 & 7.629 & 5.11\tabularnewline
\hline 
\end{tabular}
\end{table*}

To evaluate the thermodynamic stability and determine the range of chemical potentials favorable for the formation of the target compounds, we computed the Gibbs free energies of LaMoN$_{3}$ target phase at ten different pressures as well as for associated seven secondary phases using DFPT and VaspGibbs\cite{therrien2023vaspgibbs}. By employing this methodology, one can identify the specific chemical conditions under which pressure-induced LaMoN$_{3}$ systems can be synthesized. The extent of the stability region in the chemical potential space serves as an indicator of the feasibility of synthesizing the desired compound, by negating the formation of competing secondary phases. All possible secondary phases included in the analysis viz. La, Mo, N$_{2}$, LaN, LaN$_{8}$, MoN, Mo$_{3}$N$_{5}$, Mo$_{2}$N, Mo$_{3}$N$_{2}$, and Mo$_{5}$N$_{6}$ were obtained from the ICSD materials database. To determine the region of thermodynamic stability, constraint equations corresponding to each target phase and its competing secondary phases were systematically plotted on their respective chemical phase diagrams. The extent of each stability region is observed to vary as a function of pressure, with the 0 GPa phase exhibiting the most pronounced and larger stability region (Fig. \ref{fig:1}c). However, as pressure increases to 40 GPa, the stability domain for pressure-induced LaMoN$_{3}$ phase keeps on reducing, including a substantially narrow range of chemical potentials at higher pressures (Fig. S3). This pressure-dependent behavior underscores the sensitivity of phase equilibria to external thermodynamic variables. Extended discussions on the thermodynamic chemical stability can be found in our previous works\cite{padelkar2024mixed,nabi2024lead}.

\textcolor{black}{Apart from phonon and thermodynamical stability, there can be other pressure-induced effects that can impact stability. Recent studies \cite{PhysRevB.111.094108,racioppi2025activation} have demonstrated that, under sufficiently high pressure, semicore electrons can participate in bonding and influence structural phase stability. Such effects may become relevant in regimes where core-core overlap and core-level hybridization are significant. In the present work, we have not included these effects, as our calculations employ standard pseudopotentials without explicit treatment of semicore states as valence. }

\subsection{Relevance of Materials under pressure}
It is important to note that pressure is used as a design variable and not as a device operating condition in this study. 
\textcolor{black}{High pressure is a well-established computational strategy to: (i) identify hidden phases with desirable optoelectronic properties (ii) mimic chemical or epitaxial strain (iii) access electronic configurations that may later be stabilized via alloying, strain, or epitaxy (without actual high-pressure operation).}
The operating condition of any optoelectronic device indeed requires effective atmospheric pressure.
In a nutshell, any pressure-dependent studies correspond to lattice compression, which is a fundamental mechanism that can be accessed experimentally via different routes such as epitaxial strain, chemical substitution, high-entropy alloying, growth on lattice-mismatched substrates, defect engineering etc. Pressure-dependent studies on photovoltaic materials have been widely conducted to understand how their structural and electronic properties change under varying pressure.\cite{lu2023high,oyelade2020pressure}

\textcolor{black}{During compression from 0 GPa to 40 GPa, anisotropic strains of approximately ~4\% along the a- and b-axes and ~7\% along the c-axis are induced in LaMoN$_3$. Such anisotropic strain can be replicated at ambient conditions by combining various non-pressure techniques that independently modify each crystallographic direction. For instance, epitaxial growth on lattice-mismatched substrates can impose in-plane strain along the a- and b-axes, while nanostructuring methods like Vertically Aligned Nanocomposites (VAN) and nanopillar embedding or thermal quenching as a part of post-treatments, can selectively adjust the out-of-plane c-axis lattice parameter. This coordinated approach, although complex, allows for a controlled, direction-specific lattice distortions without external pressure, thus enabling the stabilization of the desired LaMoN$_3$ phase at room temperature and ambient pressure\cite{sun2018three,cao2021vertical,badding1995high}. In summary, the ability to create metastable high-pressure phases via different strategies is an important factor towards realising efficient optoelectronic properties.}

\subsection{Electronic properties}
After confirming the stability of LaMoN$_{3}$ at various pressures,
we performed electronic structure calculations, as these are crucial
for designing photoelectric devices. Electronic density
of states (DOS) and the band structures help to gain deeper insights into the electronic properties.
Initially, electronic structure calculations  under
various pressures were performed using the semilocal PBE exchange-correlation
functional, with and without SOC. However, PBE
is known to underestimate bandgaps due to self-interaction errors,
while SOC intends to reduce the bandgap (at max 0.05 eV in our present case). Consequently,
to obtain more accurate bandgap values of this system at various pressures,
we employed the hybrid HSE06 xc functional together with the MBPT-based
G$_{0}$W$_{0}$@PBE approach. The pressure-dependent evolution of
the band structures obtained from G$_{0}$W$_{0}$@PBE is displayed
in Fig. S5, covering the range from 0 to 40 GPa.
The results reveal that LaMoN$_{3}$ consistently maintains an indirect
bandgap character under all investigated pressures, as the conduction
band minimum (CBM) and valence band maximum (VBM) are situated at
different k-points in the Brillouin zone. In all the cases, the VBM is located
at the F point, while the CBM lies between the $\Gamma$ and L points;
however, with increasing pressure, the CBM gradually shifts between
the $\Gamma$ and X points. This trend suggests that compression influences
the orbital interactions and hybridization responsible for the conduction
band edge, while the valence band remains comparatively rigid. Quantitatively,
the bandgap of LaMoN$_{3}$ exhibits a monotonic decrease with increasing
pressure, as summarized in Table \ref{tab:2}. Within the HSE06 framework,
the bandgap reduces from 1.95 eV at 0 GPa to 1.39 eV at 40 GPa, whereas
the G$_{0}$W$_{0}$@PBE approach yields slightly larger values, with a decreasing trend
from 2.17 eV at ambient conditions to 1.45 eV at 40 GPa. In addition,
the direct bandgaps ($E_{g}^{dir}$) of LaMoN$_{3}$ calculated within G$_{0}$W$_{0}$@PBE framework
at different applied pressures are listed in Table \ref{tab:2}. It
is observed that the energy difference between the direct and indirect
bandgaps progressively decreases with increasing pressure, indicating
that LaMoN$_{3}$ tends toward a more direct-gap character under compression.
This reduction in the direct-indirect gap separation is particularly
significant for optoelectronic applications, as it enhances the likelihood
of direct optical transitions and improves light absorption efficiency.

\begin{figure*}[t]  
\centering
\includegraphics[width=1.0\linewidth]{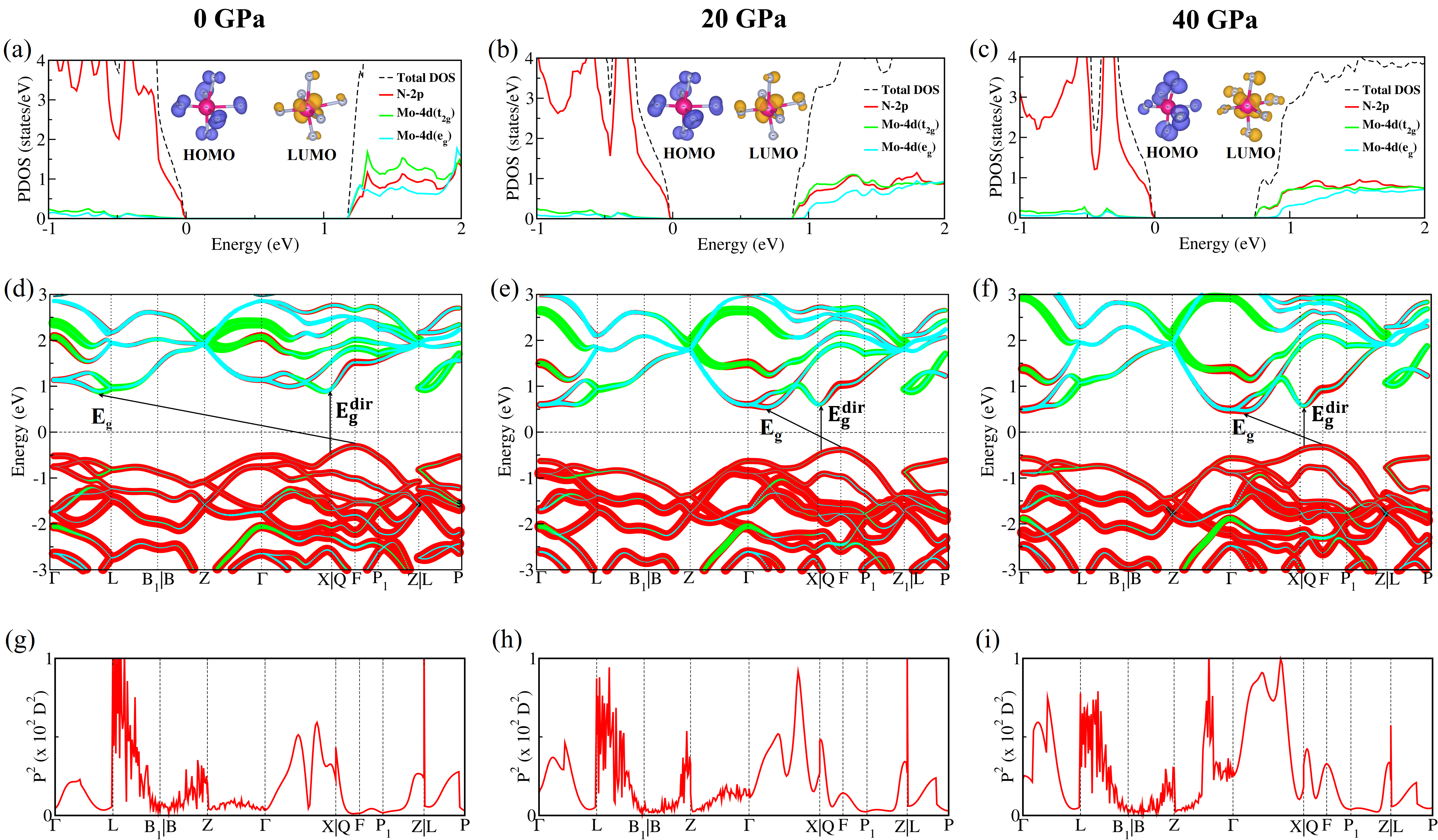}
\centering
\caption{Electronic structure of LaMoN$_{3}$ under various pressures: (a-c) Orbital-projected density of states (PDOS) simulated using PBE functional at 0 GPa, 20 GPa, and 40 GPa pressure. The inset in each of the  panel demonstrates the isosurface plots of band decomposed partial charge density at HOMO and LUMO. Here, the pink and silver spheres represent Mo and N atoms, respectively, (d-f) Orbital-projected band structure at 0 GPa, 20 GPa, and 40 GPa pressure. The color scheme of band projection remains the same as that of PDOS. ${E}_{g}$ and ${E}_{g}^{{dir}}$ are the electronic indirect and smallest direct bandgaps, and (g-i)  Transition probability (square of the transition dipole moment matrix elements) for 0 GPa, 20 GPa, and 40 GPa. The unit of the transition dipole moment is Debye (D).} 
\label{Figure-2}
\end{figure*}

The observed reduction in bandgap with increasing pressure from 0 to 40 GPa is predominantly driven by the downward shift of the conduction bands. To gain further understanding of the mechanisms behind this behavior, we simulated the orbital-projected density of states and electronic band structure of three representative case, 0 GPa, 20 GPa, and 40 GPa LaMoN$_{3}$ structures as indicated in Fig. \ref{Figure-2}(a-f). The effect of pressure on band gap variation can be interpreted using crystal field theory and the orbital projections near the band edge states.
The electronic structure of LaMoN$_{3}$ is mainly driven by the interactions between Mo-4d and N-2p orbitals within an octahedral crystal field. At 0 GPa (Fig. \ref{Figure-2}(a,d)), the HOMO is primarily composed of N-2p orbitals, whereas the LUMO is dominated by Mo-4d states. Such orbital differentiation at the band edges is a direct consequence of crystal field splitting and covalent contributions arising from the distinct local environments around Mo and N. The  crystal field splitting causes Mo 4d orbitals to separate into lower-energy t$_{2g}$ and higher-energy e$_g$ levels, with the Mo-4d (t$_{2g}$) forming the lowest-lying conduction states due to the charge density lobes lying in the x-y orientation, as observed in the partial charge density decoration at LUMO (Fig. \ref{Figure-2}a(inset)). Meanwhile, the N-2p orbitals, located at the apex of the valence band (HOMO), mainly maintain ligand-derived character as a result of weaker hybridization with Mo-4d orbitals. This pronounced difference in symmetry and spatial character between the HOMO and LUMO suppresses direct orbital overlap, thereby increasing the transition energy required for electronic excitation, while keeping the band gap relatively wide.

Upon application of external pressure, the contraction of the unit cell and the shortening of Mo--N bond lengths intensify the orbital interactions. Crystal field theory predicts a simultaneous enhancement of covalent character and orbital hybridization between the Mo-4d and N-2p states. Notably, this pressure-induced evolution in electronic structure manifests as a pronounced shift in the character of both the HOMO and LUMO, exhibiting dominant N-2p orbital contributions. At 20 GPa (Fig. \ref{Figure-2}(b,e)), increase in presence of N-2p states is observed around the LUMO, as evidenced by PDOS, band structure projections, and partial charge density analyses. This hybridization is further intensified at 40 GPa (Fig. \ref{Figure-2}(c,f)), where the LUMO charge density lobes show a marked increase in spatial extension relative to the ambient pressure condition. Conversely, the partial charge density distribution at the HOMO remains largely invariant with respect to applied pressure. The enhanced dominance of N-2p orbitals within the conduction band minimum induces a downward energy shift in the conduction bands. The spatial alignment of these orbitals under high-pressure  facilitates substantial orbital overlap, thereby promoting more efficient electronic transitions. This reconfiguration consequently results in a significant reduction in the bandgap, arising from the diminished energy separation between the HOMO and the LUMO levels, due to increased Mo--N hybridization and redistribution of the electron charge density within the octahedral crystal field.

To deepen our understanding of pressure induced charge
carrier transport further, we calculated the effective
masses of electrons ($m_{e}^{*}$) and holes ($m_{h}^{*}$) at various
pressures and the corresponding values are presented in Table \ref{tab:2}.
The effective mass
serves as a crucial parameter for evaluating carrier mobility and
hence directly influences the overall optoelectronic performance of
the material. The value of $m_{e}^{*}$ and $m_{h}^{*}$ in LaMoN$_{3}$
are found to increase with applied pressure, with the exception that
$m_{h}^{*}$ decreases slightly up to 5 GPa before increasing again
at higher pressures. The computed values fall within the range of
0.738-1.493 for electrons and 1.000-7.629 for holes. This significant
disparity indicates that the electron mobility in LaMoN$_{3}$ remains
always higher than the hole mobility. Such differences in carrier
transport play an important role in governing recombination dynamics.
The variation in mobilities reduces the likelihood of electron-hole
recombination, thereby favoring efficient charge separation. To quantify
this effect, the relative effective mass ratio ($M)$ is introduced
as \citep{Ref.38}, $M=\left|m_{h}^{*}\right|/\left|m_{e}^{*}\right|$,
where a larger $M$ value reflects a greater asymmetry between electron
and hole mobilities. A higher ratio thus corresponds to a reduced
recombination rate of electron-hole pairs, which is beneficial for
optoelectronic performance. As summarized in Table \ref{tab:2}, the
calculated $M$ values increase steadily with pressure beyond 10 GPa,
ranging from 1.30 to 5.11. This trend clearly suggests that external
compression should help in suppressing the recombination processes and improving the optoelectronic
efficiency of LaMoN$_{3}$.

\textcolor{black}{The pressure-induced evolution of structural parameters provides a direct microscopic origin for the observed electronic and optical trends. As pressure increases, the systematic reduction in Mo-N bond length, bond angle variance, distortion index, and octahedral volume indicates progressive symmetrization of the MoN$_6$ octahedra and reduced cation off-centering. This structural compression enhances Mo-4d - N-2p orbital overlap, leading to stronger hybridization and increased band dispersion. Consequently, the conduction band shifts downward, resulting in the observed reduction in bandgap. At the same time, the enhanced orbital overlap increases the electronic polarizability of the system, which manifests as an increase in the dielectric constant and plays a crucial role in modifying excitonic and optical properties.}

\subsection{Optical Properties and Spectroscopic Limited Maximum Efficiency}

To evaluate the suitability of a material for 
optoelectronic applications, it is essential to conduct a detailed
analysis of its optical properties, particularly the dielectric function
and absorption coefficient. To enhance predictive accuracy, we employed
MBPT within the GW-BSE framework to
calculate the optical response, explicitly incorporating electron-hole
interactions \citep{Ref.31,Ref.32}. This was done by first performing a single-shot
GW (G$_{0}$W$_{0}$) calculations based on the PBE functional. The
resulting quasiparticle energies were then used to solve the BSE. The frequency-dependent dielectric function, $\varepsilon(\omega)$,
was subsequently computed as $\varepsilon(\omega)$ = {[}Re($\varepsilon$){]}
+ i{[}Im($\varepsilon$){]}, where {[}Re($\varepsilon$){]} and {[}Im($\varepsilon$){]}
correspond to the real and imaginary parts, respectively.

\begin{figure*}[htbp]  
\centering
\includegraphics[width=0.6\linewidth]{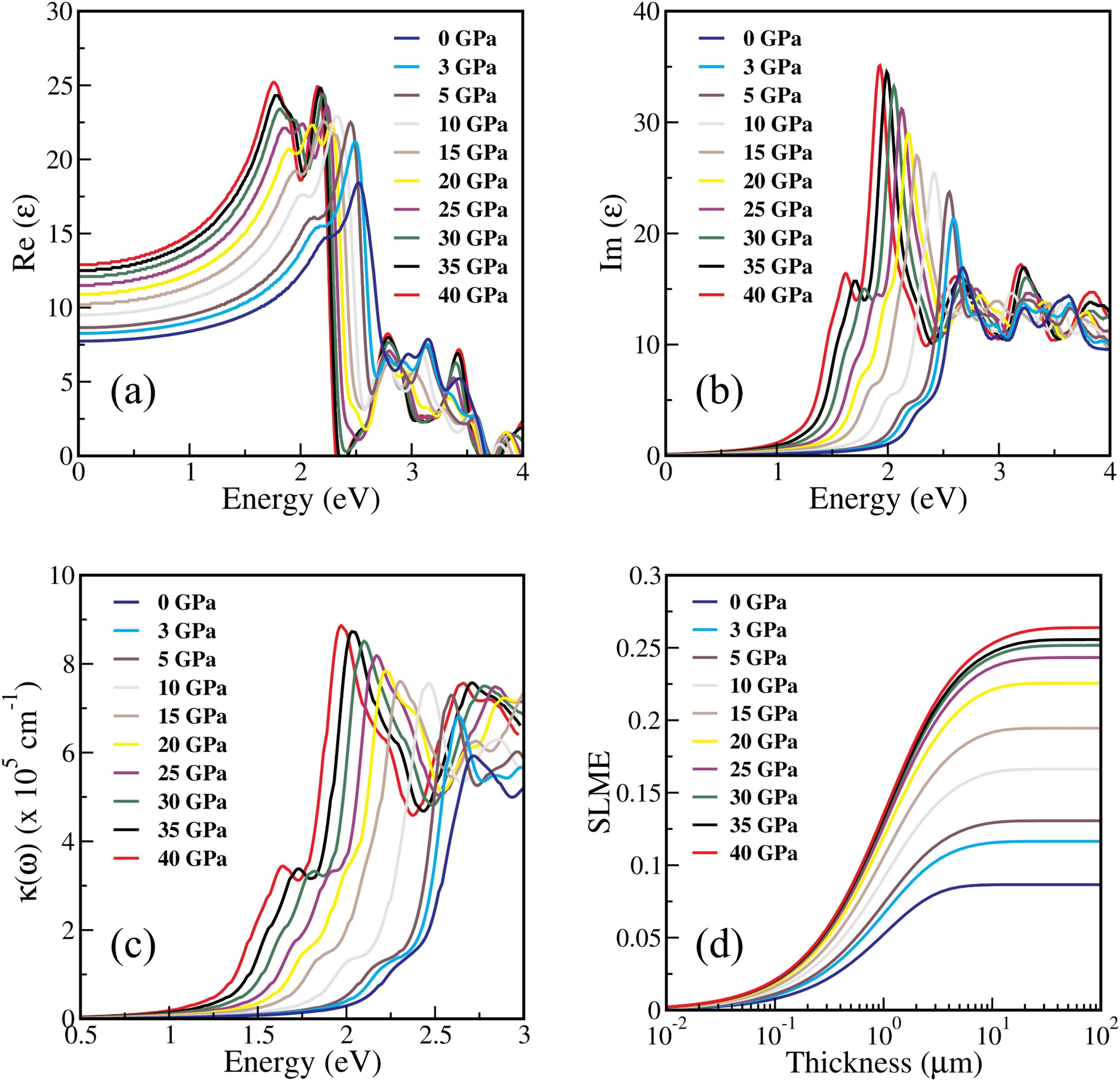}
\centering
\caption{\label{fig:3}(a) Real and (b) imaginary part of the electronic dielectric function ($\varepsilon$),  (c) \textcolor{black}{absorption coefficient, $\kappa(\omega)$,} and (d) SLME of LaMoN$_{3}$ at various pressures calculated using the BSE@G$_{0}$W$_{0}$@PBE method.} 
\end{figure*}

The real part of the dielectric function, {[}Re($\varepsilon$){]},
reflects the material\textquoteright s polarization response to an
external electric field. A higher value of Re($\varepsilon$) corresponds
to stronger polarization, which directly affects key optical properties
such as the refractive index and light absorption. The calculated
{[}Re($\varepsilon$){]} of LaMoN$_{3}$ at various pressures, obtained
using BSE@G$_{0}$W$_{0}$@PBE, is presented in Fig. \ref{fig:3}(a).
 Re($\varepsilon$) at zero energy defines the electronic
or optical dielectric constant ($\varepsilon_{\infty}$), which describes
the dielectric screening of electron-hole Coulomb interactions. Our
results demonstrate that $\varepsilon_{\infty}$ increases with applied
pressure from 0 GPa to 40 GPa. This pressure-induced enhancement suggests
reduced charge-carrier recombination and improved optoelectronic performance,
as a result of stronger electronic
screening and a more robust dielectric response. This observation is consistent with the effective mass ratio predictions.

On the other hand, the imaginary part of the dielectric function,
{[}Im($\varepsilon$){]}, is crucial for describing the linear absorption
properties of a material. As shown in Fig. \ref{fig:3}(b), both
the absorption edge and the position of the first peak ($E_{o}$)
of LaMoN$_{3}$ gradually red-shift (from 2.25 eV to 1.51 eV) with increasing pressure from
0 GPa to 40 GPa. This red shift directly correlates with the reduction
in the quasiparticle (QP) bandgap, as summarized in Table \ref{tab:2}.
 Notably, the absorption edge of LaMoN$_{3}$ reduces
from the near-visible to the infrared region with increasing pressure,
which may enhance the achievable power conversion efficiency (PCE)
and make LaMoN$_{3}$ a promising candidate for high-performance photovoltaic
applications under pressure.

One of the most important parameters for assessing the suitability
of a material in optoelectronic applications is the absorption coefficient,
which directly reflects its capability for efficient solar energy
conversion. To evaluate this property, the absorption coefficient,
{[}$\kappa(\omega)${]}, for the investigated perovskite is calculated
using the following equation\citep{Ref.20}:

\begin{equation}
\kappa(\omega)=\sqrt{2}\omega\text{\ensuremath{\left[\sqrt{\{\mathrm{Re}(\varepsilon)\}^{2}+\{\mathrm{Im}(\varepsilon)\}^{2}}-\mathrm{Re}(\varepsilon)\right]^{1/2}.}}
\end{equation}

Fig. \ref{fig:3}(c) shows our BSE@G$_{0}$W$_{0}$@PBE results for $\kappa(\omega)$ of LaMoN$_{3}$ under varying pressure, exhibiting a high absorption coefficient on the
order of 10$^{5}$ cm$^{-1}$ across all applied pressures, highlighting
its potential for optoelectronic applications. To further assess the optoelectronic performance of LaMoN$_{3}$, we have calculated the SLME, as proposed by Yu and Zunger \citep{Ref.39} (for
details, see the SI). SLME is an improved performance
metric over the conventional Shockley-Queisser (SQ) efficiency limit
\citep{Ref.40}, specifically developed to estimate the theoretical
maximum efficiency of thin-film absorber materials. Unlike the oversimplified SQ
limit, which neglects the losses from radiative
recombination due to non-conservation of photon momentum, the SLME
model incorporates additional material-specific factors. These include
the magnitude and nature of the bandgap (direct or indirect), the
detailed shape of the absorption spectrum, absorber layer thickness,
non-radiative recombination losses, and temperature. In this work,
the SLME of the investigated perovskite is evaluated at 293.15 K under
the standard solar spectrum (AM1.5G), using the absorption coefficient,
material thickness, and the electronic bandgap obtained from G$_{0}$W$_{0}$@PBE
calculations.

To perform the SLME calculation, we first investigated the optical
transition possibilities at the lowest direct band edges. In many
materials, such transitions may be forbidden due to inversion symmetry
in the crystal structure, which imposes identical parity on the direct
band edge states. To verify whether such transitions are optically
allowed in our case, we compute the transition dipole moment matrix
element ($\mathrm{P}$), the squared modulus ($\mathrm{P}^{2}$) of which
directly quantifies the transition probability. The calculated $\mathrm{P}^{2}$
values for LaMoN$_{3}$ at 0 GPa, 20 GPa, and 40 GPa are plotted in
Fig. \ref {Figure-2}(g,h,i). These results clearly indicate that LaMoN$_{3}$
exhibits allowed dipole transitions at the lowest direct band edges.
However, as LaMoN$_{3}$ is an indirect bandgap material, non-radiative
recombination plays a dominant role in influencing the SLME. In contrast,
radiative recombination is corrected by a factor, $f_{r}=e^{(E_{g}-E_{g}^{da})/k_{B}T}$, where $E_{g}$ is the fundamental electronic bandgap, $E_{g}^{da}$ is the direct
allowed bandgap, $k_{B}$ is the Boltzmann constant, and $T$ is the
temperature \citep{Ref.39} (further details are provided in the SI). \textcolor{black}{The indirect nature of the band gap plays a critical role in determining photovoltaic performance. In indirect semiconductors, the absorption onset is governed by phonon-assisted transitions, resulting in a reduced absorption coefficient near the band edge compared to direct-gap materials. Within the SLME formalism, this effect is inherently captured through the absorption spectrum $\kappa(\omega)$, which directly influences the short-circuit current density ($J_{sc}$). Additionally, the difference between the fundamental and direct allowed band gaps ($E_{g}^{da}-E_{g}$) is explicitly incorporated via the radiative recombination factor $f_{r}$, such that indirect band gap materials ($E_{g}^{da} > E_{g}$) exhibit enhanced nonradiative recombination losses and reduced open-circuit voltage.}

After confirming the presence of optically allowed dipole transitions
at the lowest direct band edges, thickness-dependent
SLME calculations are performed for all applied pressures using the
BSE@G$_{0}$W$_{0}$@PBE results. As shown in Figure \ref{fig:3}(d), the SLME of LaMoN$_{3}$ exhibits a sharp increase with increasing applied
pressure. Specifically, the SLME rises from 8.66\% at 0 GPa to 26.38\%
at 40 GPa, highlighting the significant enhancement of its photovoltaic
performance under compression. \textcolor{black}{Under applied pressure, both the fundamental and direct band gaps of LaMoN$_3$ decrease within an optimal pressure region, accompanied by a reduction in ($E_{g}^{da}-E_{g}$). This leads to improved absorption characteristics, an increase in $J_{sc}$, and an enhanced radiative recombination fraction. Consequently, the system exhibits a transition toward direct-gap-like behavior, resulting in a significant enhancement in the predicted SLME.} These findings suggest that LaMoN$_{3}$
possesses promising optoelectronic potential at high pressures, making
it a viable candidate for pressure-tuned thin-film solar cell applications.

\subsection{Excitonic Properties}

Beyond the optical response, we also examine key excitonic parameters,
namely the exciton binding energy ($E_{B}$) and exciton temperature
($T_{exc}$), which play a central role in determining the efficiency
of photoexcited charge dynamics in PV materials. When light is absorbed,
an electron ($e$) and a hole ($h$) may remain bound together as
an exciton. The stability of such excitons has a direct impact on
optoelectronic performance: while weakly bound excitons readily dissociate
into free carriers that contribute to current, strongly bound excitons
can be advantageous for other optoelectronic applications.

The exciton binding energy ($E_{B}$) quantifies the energy cost required
to separate the electron-hole pair. Within first-principles theory,
$E_{B}$ is extracted as the difference between the direct quasiparticle
(G$_{0}$W$_{0}$@PBE) bandgap, $E_{g}^{dir}$, and the optical bandgap,
$E_{o}$ (position of the first peak of BSE@G$_{0}$W$_{0}$@PBE simulated spectra)
\citep{Ref.43,Ref.42}. The computed $E_{B}$ values
of LaMoN$_{3}$ under different applied pressures are summarized in
Table \ref{tab:3}. A monotonic decrease in $E_{B}$ is observed with
increasing pressure. At 0 GPa, $E_{B}$ is relatively high, around
216 meV, but it gradually decreases to 64 meV at 40 GPa. This result
indicates that applying external pressure effectively reduces the
$E_{B}$ of LaMoN$_{3}$, bringing it close to the range reported
for conventional lead-based halide perovskites (10$-$100 meV) \citep{Ref.45,Ref.46,Ref.47}.
The ability to access both low- and high-$E_{B}$ regimes through
pressure modulation makes LaMoN$_{3}$ a versatile candidate for a
wide range of optoelectronic applications.

Another important parameter is the excitonic temperature ($T_{exc}$),
which represents the maximum temperature at which an exciton remains
stable. It is related to the thermal energy required for exciton dissociation
through the relation $E_{B}=k_{B}T_{exc}$, where $k_{B}$ is the
Boltzmann constant. As shown in Table \ref{tab:3}, excitons in LaMoN$_{3}$
are stable up to about 2504 K at 0 GPa, whereas under 40 GPa pressure
they remain stable only up to 742 K.

\begin{table*}
\caption{\label{tab:3}Calculated excitonic parameters\textemdash including
exciton binding energy ($E_{B}$), excitonic temperature ($T_{exc}$),
characteristic phonon angular frequency ($\omega_{LO}$), electronic
($\varepsilon_{\infty}$) and static ($\varepsilon_{s}$) dielectric
constants, phonon screening correction ($\Delta E_{B}^{ph}$), and
modified exciton binding energy ($E_{B}^{\prime}$) of LaMoN$_{3}$
 under different applied pressures.}

\centering{}%
\begin{tabular}{cccccccc}
\hline 
\multirow{1}{*}{Pressure (GPa)} & $E_{B}$ (meV) & $T_{exc}$ (K) & $\omega_{LO}$ (THz) & $\varepsilon_{\infty}$ & $\varepsilon_{s}$ & $\Delta E_{B}^{ph}$ (meV) & $E_{B}^{\prime}$ (meV)\tabularnewline
\hline 
0 & 216 & 2504 & 9.78 & 7.74 & 66.64 & -32.09 & 184\tabularnewline
3 & 136 & 1577 & 9.88 & 8.27 & 67.80 & -30.33 & 106\tabularnewline
5 & 108 & 1252 & 9.89 & 8.65 & 69.73 & -29.13 & 79\tabularnewline
10 & 97 & 1125 & 10.00 & 9.50 & 72.34 & -28.55 & 68\tabularnewline
15 & 85 & 985 & 10.11 & 10.20 & 73.44 & -27.76 & 57\tabularnewline
20 & 78 & 904 & 10.43 & 10.90 & 74.42 & -27.62 & 50\tabularnewline
25 & 73 & 846 & 10.82 & 11.50 & 75.14 & -27.70 & 45\tabularnewline
30 & 70 & 812 & 11.02 & 12.10 & 75.24 & -27.50 & 42\tabularnewline
35 & 67 & 777 & 11.16 & 12.50 & 74.42 & -27.18 & 40\tabularnewline
40 & 64 & 742 & 11.31 & 12.90 & 74.02 & -26.87 & 37\tabularnewline
\hline 
\end{tabular}
\end{table*}

In the present study, $E_{B}$ is calculated using the standard ab-initio
BSE approach. However, a limitation of this method is that it considers
only electronic screening when constructing the $e$$-$$h$ kernel.
This static screening does not account for the electron-phonon
coupling, which can be crucial in polar materials, especially in systems
with significant electron-phonon interactions or where phonons play
a key role in determining optoelectronic properties. In a recent study,
Filip et al. \citep{Ref.48} incorporated phonon screening into the
calculation of $E_{B}$ by assuming isotropic and parabolic electronic
band dispersion. The phonon-screening correction ($\Delta E_{B}^{ph}$)
is expressed as,

\[
\Delta E_{B}^{ph}=-2\omega_{LO}\left(1-\frac{\varepsilon_{\infty}}{\varepsilon_{s}}\right)\frac{\sqrt{1+\omega_{LO}/E_{B}}+3}{\left(1+\sqrt{1+\omega_{LO}/E_{B}}\right)^{3}},
\]

where $\omega_{LO}$ denotes the characteristic phonon angular frequency,
while $\varepsilon_{\infty}$ and $\varepsilon_{s}$ represent the
electronic (optical) and static dielectric constants, respectively.
The static dielectric constant ($\varepsilon_{s}$) is defined as
$\varepsilon_{s}=$$\varepsilon_{\infty}+\varepsilon_{ion}$, where
$\varepsilon_{ion}$ is the ionic dielectric constant obtained from
DFPT calculation. To evaluate $\omega_{LO}$, we employ the thermal
\textquotedbl B\textquotedbl{} approach developed by Hellwarth et
al. \citep{Ref.44}, which provides an effective average by accounting
for the spectral contributions of multiple phonon branches (for details,
see SI). The contribution of phonon screening
is found to reduce $E_{B}$ by 26.87$-$32.09 meV (see Table \ref{tab:3}), which is significant
 and therefore cannot be neglected in the present case. After incorporating
this correction, the modified exciton binding energy ($E_{B}^{\prime}$)
of LaMoN$_{3}$ is 184 meV at 0 GPa and decreases to 37 meV at 40
GPa, underscoring the pressure-induced tunability of this material
and its potential as a promising candidate for optoelectronic applications.

\textcolor{black}{The observed decrease in exciton binding energy with pressure can be directly correlated with the enhancement in dielectric screening. As the electronic dielectric constant increases from ~7.7 to ~12.9 with increasing pressure, the Coulomb interaction between electrons and holes is effectively screened. This reduction in electron–hole interaction strength lowers the exciton binding energy from 216 meV to 64 meV, facilitating easier charge separation and improving photovoltaic performance.}

\subsection{Polaronic Properties}

To understand the fundamental limits of carrier mobility in polar
semiconductors, it is essential to account for the interaction between
charge carriers and longitudinal optical (LO) phonons. In polar materials,
the dominant scattering mechanism near room temperature originates
from this electron (or hole)-phonon coupling, which gives rise to
a polaron state rather than a purely free-carrier state. This carrier-phonon
interaction can be effectively captured within the Fr\"ohlich model, where
the coupling strength is quantified by the dimensionless Fr\"ohlich
parameter, $\alpha$, defined as\citep{Ref.9,Ref.36,Ref.Adhikari},

\begin{equation}
\alpha=\frac{1}{4\pi\varepsilon_{0}}\frac{1}{2}\Big(\frac{1}{\varepsilon_{\infty}}-\frac{1}{\varepsilon_{s}}\Big)\frac{e^{2}}{\hbar\omega_{LO}}\Big(\frac{2m^{\ast}\omega_{LO}}{\hbar}\Big)^{1/2}
\end{equation}

where $\varepsilon_{0}$ represents the permittivity of free space
and $m^{\ast}$ is the carrier effective mass. Table \ref{tab:4}
shows that the computed values of $\alpha$ for electrons and holes
in LaMoN$_{3}$ under different applied pressures lie within the intermediate
coupling regime (1.33$-$3.01), where $\alpha\ll1$ corresponds to
weak coupling and $\alpha>10$ indicates strong coupling \citep{Ref.41}.
The relatively weaker electron-phonon coupling, which decreases with
increasing pressure, can be attributed to the smaller electron effective
mass and the larger electronic dielectric constant. In contrast, the
comparatively stronger hole-phonon coupling, which exhibits an oscillatory
dependence on pressure, originates from the relatively heavier effective mass
of holes.

In polar material, charge carriers (electrons
or holes) interact with the lattice, inducing distortions in the surrounding
ions and leading to polaron formation. This process lowers the QP
energy, as both electrons and holes lose energy upon forming polarons.
The corresponding polaron energy ($E_{p}$) is derived from the coupling
constant ($\alpha$) and calculated using the following expression
\citep{Ref.9,Ref.36}:
\[
E_{p}=(-\alpha-0.0123\alpha^{2})\hbar\omega_{LO}.
\]

The QP gap of LaMoN$_{3}$ under different applied pressures, obtained from the combined polaron energies of electrons and holes (see Table \ref{tab:4}),
is computed and compared with $E_{B}$ from Table \ref{tab:3}. For
LaMoN$_{3}$, the comparison shows that the energy of charge-separated
polaronic states remains higher than that of bound exciton states
at all applied external pressures except at 0 GPa. This indicates
that charge-separated polaronic states (where carriers are spatially
separated) are more stable than bound excitons (where electrons and
holes remain closely associated) under pressure, while at 0 GPa the
opposite trend is observed, with bound excitonic states being energetically
favored.

In polarons, the carrier\textquoteright s effective mass increases
due to its interaction with lattice vibrations (phonons), as the surrounding
phonon cloud enhances its mass and alters its mobility. Therefore,
the effective mass of the polaron ($m_{p}$) is defined using the
Feynman's extended version of Fr\"ohlich's polaron theory (for a small
$\alpha$), which can be expressed as follows \citep{Ref.35,Ref.42}:
\[
m_{p}=m^{\ast}\Big(1+\frac{\alpha}{6}+\frac{\alpha^{2}}{40}+...\Big)
\]

where $m^{\ast}$ is the carrier effective mass. Table \ref{tab:4}
shows that carrier-phonon coupling increases the polaron effective
mass of LaMoN$_{3}$ by 27$-$73\% under varying pressures, indicating
intermediate carrier-lattice interactions.

Finally, the polaron mobility ($\mu_{p}$), which captures the effect
of the increased polaron effective mass on carrier transport, is evaluated
using the Hellwarth polaron model \citep{Ref.36,Ref.44} as:
\[
\mu_{p}=\frac{\left(3\sqrt{\pi}e\right)}{\omega_{LO}m^{*}\alpha}\frac{\sinh(\beta/2)}{\beta^{5/2}}\frac{w^{3}}{v^{3}}\frac{1}{K(a,b)}
\]
where $e$ is the electronic charge, $\beta=\hbar\omega_{LO}/k_{B}T$,
$w$ and $v$ are the temperature-dependent variational parameters,
and $K(a,b)$ is a function of $\beta$, $w$, and $v$ (for details,
see the SI). As intermediate carrier-phonon
coupling raises the effective mass, mobility generally decreases,
slowing polaron motion, which is shown in Table \ref{tab:4}. Our
results show that the polaron mobility of electrons (15.08$-$21.78
cm$^{2}$V$^{-1}$s$^{-1}$) in LaMoN$_{3}$ under applied pressures
is consistently higher than that of holes (0.86$-$13.96 cm$^{2}$V$^{-1}$s$^{-1}$),
primarily due to the influence of the polaron effective mass. Overall,
our study indicates that pressure strongly affects the polaronic properties
of LaMoN$_{3}$, which plays a key role in determining its pressure-dependent
optoelectronic performance.

\begin{table*}
\caption{\label{tab:4}Calculated polaron parameters\textemdash including Fr\"ohlich
interaction parameter ($\alpha$), polaron energy ($E_{p}$), polaron effective mass ($m_{p}$), and polaron mobility ($\mu_{p}$)
for electrons ($e$) and holes ($h$) of LaMoN$_{3}$ under different
applied pressures.}

\centering{}%
\begin{tabular}{ccccccccccccc}
\hline 
\multirow{2}{*}{Pressure (GPa)} &  & \multicolumn{2}{c}{$\alpha$} &  & \multicolumn{2}{c}{$E_{p}$ (meV)} &  & \multicolumn{2}{c}{$m_{p}/m^{*}$} &  & \multicolumn{2}{c}{$\mu_{p}$ (cm$^{2}$V$^{-1}$s$^{-1}$)}\tabularnewline
\cline{3-4}\cline{6-7}\cline{9-10}\cline{12-13}
 &  & $e$  & $h$ &  & $e$  & $h$ &  & $e$  & $h$ &  & $e$  & $h$\tabularnewline
\hline 
0 &  & 1.80 & 2.31 &  & 74.52 & 96.22 &  & 1.38 & 1.52 &  & 19.63 & 8.24\tabularnewline
3 &  & 1.69 & 2.10 &  & 70.58 & 88.14 &  & 1.35 & 1.46 &  & 20.84 & 9.84\tabularnewline
5 &  & 1.62 & 1.85 &  & 67.67 & 77.49 &  & 1.34 & 1.39 &  & 21.78 & 13.96\tabularnewline
10 &  & 1.55 & 2.00 &  & 65.41 & 84.86 &  & 1.32 & 1.43 &  & 20.52 & 8.53\tabularnewline
15 &  & 1.56 & 1.98 &  & 66.57 & 84.92 &  & 1.32 & 1.43 &  & 16.86 & 7.47\tabularnewline
20 &  & 1.45 & 1.99 &  & 63.75 & 88.06 &  & 1.29 & 1.43 &  & 18.06 & 6.19\tabularnewline
25 &  & 1.38 & 1.84 &  & 62.88 & 84.31 &  & 1.28 & 1.39 &  & 18.40 & 6.93\tabularnewline
30 &  & 1.36 & 2.11 &  & 63.10 & 98.79 &  & 1.27 & 1.46 &  & 16.87 & 3.79\tabularnewline
35 &  & 1.36 & 2.40 &  & 63.91 & 114.19 &  & 1.27 & 1.54 &  & 15.43 & 2.18\tabularnewline
40 &  & 1.33 & 3.01 &  & 63.31 & 146.20 &  & 1.27 & 1.73 &  & 15.08 & 0.86\tabularnewline
\hline 
\end{tabular}
\end{table*}

\textcolor{black}{Our results reveal a clear interplay between optical enhancement and carrier transport under applied pressure. While pressure significantly improves the optical absorption and SLME of LaMoN$_{3}$, it also strengthens carrier-phonon coupling, leading to an increase in polaron effective mass and a corresponding reduction in mobility. However, the Fröhlich coupling remains within the intermediate regime, and the resulting mobility degradation is moderate. Notably, the electron mobility remains relatively high, ensuring efficient carrier transport. At the same time, the large absorption coefficient (10$^{5}$ cm$^{-1}$) enables efficient photon absorption within thin layers, thereby reducing the dependence on long carrier diffusion lengths. Consequently, the pressure-induced enhancement in optical properties compensates for the moderate reduction in mobility. These findings indicate that the overall photovoltaic performance of LaMoN$_{3}$ improves under pressure, highlighting its potential as a pressure-tunable optoelectronic material.}

\subsection{Nonlinear Optical Properties}

The investigation of NLO properties in LaMoN$_{3}$ system is particularly compelling due to its non-centrosymmetric crystal structure, a key requirement for second-order NLO phenomena to occur. One such phenomenon, second harmonic generation (SHG), predominantly observed in piezo/ferroelectric system, involves the conversion of two photons of the same frequency into a single photon with twice the frequency, offering rich insight into the material's symmetry and electronic structure. Previous pressure-dependent studies\cite{Ref.15} on this system have reported polarization values that indicate the onset of ferroelectricity induced by structural changes under pressure. This behavior not only confirms the presence of spontaneous polarization but also implies tunability of ferroelectric properties via external pressure. These findings are especially significant for BPVE\cite{ji2011evidence}, offering an alternative paradigm to conventional p-n junction photovoltaic devices. Due to the intrinsic asymmetry and polar nature of the material, BPVE can potentially deliver photovoltages that can exceed bandgap limits and generate photocurrent without p-n junctions, thus promising higher efficiency and novel device architectures. The absence of centrosymmetry and the presence of ferroelectricity in LaMoN$_{3}$ system make it a promising candidate for BPVE, which holds the potential to surpass the PV efficiency and photovoltage limitations of conventional single junction solar cells.

Central to this prospect is the concept of shift current\cite{ibanez2018ab}, a nonlinear photocurrent arising intrinsically from the electronic structure in non-centrosymmetric materials. Shift current correlates directly with the efficiency of bulk photovoltaic conversion and thus serves as a critical metric in evaluating the material's suitability for next-generation photovoltaic technology. Despite the ferroelectric characteristics of LaMoN$_{3}$, prior studies have not thoroughly examined shift current responses under varying conditions, leaving a gap in the comprehensive understanding of this system's potential for BPVE. To address this, we have undertaken computational calculations of the shift current response, aiming to elucidate the fundamental mechanisms and quantify the material's capability to generate bulk photocurrents.

The NLO effect arises as a consequence of the response of a non-centrosymmetric material to an applied optical field with a dependence of the current density on the strength of the field  in a nonlinear manner as,\cite{sturman2021photovoltaic} 

\begin{equation}
       j_i= \sigma_{ij}^{d}E_j + \sigma_{ijk}E_j E_k + \sigma_{ijkl}^{ph}E_j p_k p_{l}^{*}I_0 + \chi_{ijkl}q_j p_k p_{l}^{*}I_0 + \beta_{ijk}p_j p_{k}^{*}I_0 + ...,
\end{equation}

where $\sigma_{ij}^{d}$ is the intrinsic linear Ohmic conductivity which is predominantly observed in conventional centrosymmetric semiconductor photovoltaic effect. For a non-centrosymmetric system, higher rank tensors arise viz., the 2nd term is the quadratic electrical conductivity $\sigma_{ijk}$, and the 5th term photo-ferroic tensor $\beta_{ijk}$, which is identical to the piezoelectric tensor. $\sigma_{ijkl}^{ph}$ and $\chi_{ijkl}$ correspond to anisotropic photoconductivity and light entrainment effect respectively. The dominant mechanism channelling the BPV effect is $\sigma_{ijk}$ and $\beta_{ijk}$, which can be related to the material dependent parameters namely shift current (and/or Ballistic current) and Glass co-efficient.

A perturbative analysis treating the electromagnetic field classically within the dipole approximation leads to the formulation of the shift current expression given by\cite{sipe2000second}

\begin{equation}
\begin{split}
\sigma_{ijk}(\omega) = \pi e \left( \frac{e}{m \hslash \omega} \right)^2
\sum_{n',n''} \int d\mathbf{k} \big[ f[n''\mathbf{k}] - f[n'\mathbf{k}] \big] \\
\times \langle n' \mathbf{k} | \hat{P}_i | n'' \mathbf{k} \rangle 
\langle n'' \mathbf{k} | \hat{P}_j | n' \mathbf{k} \rangle \\
\times \left( - \frac{\partial \phi_{n'n''}(\mathbf{k}, \mathbf{k})}{\partial k_k} 
- [ \chi_{n''k}(\mathbf{k}) - \chi_{n'k}(\mathbf{k}) ] \right) \\
\times \delta \big( \omega_{n''}(\mathbf{k}) - \omega_{n'}(\mathbf{k}) \pm \omega \big)
\end{split}
\end{equation}

\begin{equation}
J_i = \sigma_{ijk} E_j E_k
\end{equation}

where, $n'$ and $n''$ denote band indices, $\mathbf{k}$ is the wave vector within the Brillouin zone, and $\omega_n(\mathbf{k})$ represents the energy of the $n$th band. The quantity $\sigma_{ijk}$ describes the current density response $\mathbf{J}$ to the electromagnetic field $\mathbf{E}$. The Berry connections for band $n$ at $\mathbf{k}$ are given by $\chi_{nk}(\mathbf{k})$, while $\phi_{n', n''}$ indicates the phase of the transition dipole between bands $n'$ and $n''$.

LaMoN$_{3}$ belongs to space group \emph{R3c}. In this coordinate system, the form of a third-rank tensor, such as the shift current response tensor, is symmetry constrained to have the following finite components

\begin{equation}
\sigma = \begin{bmatrix}
0 & 0 & 0 & 0 & \sigma_{yyz} & -\sigma_{yyy} \\
-\sigma_{yyy} & \sigma_{yyy} & 0 & \sigma_{yyz} & 0 & 0 \\
\sigma_{zxx} & \sigma_{zxx} & \sigma_{zzz} & 0 & 0 & 0
\end{bmatrix}
\end{equation}

\begin{figure*}[h]  
\centering
\includegraphics[width=1.0\linewidth]{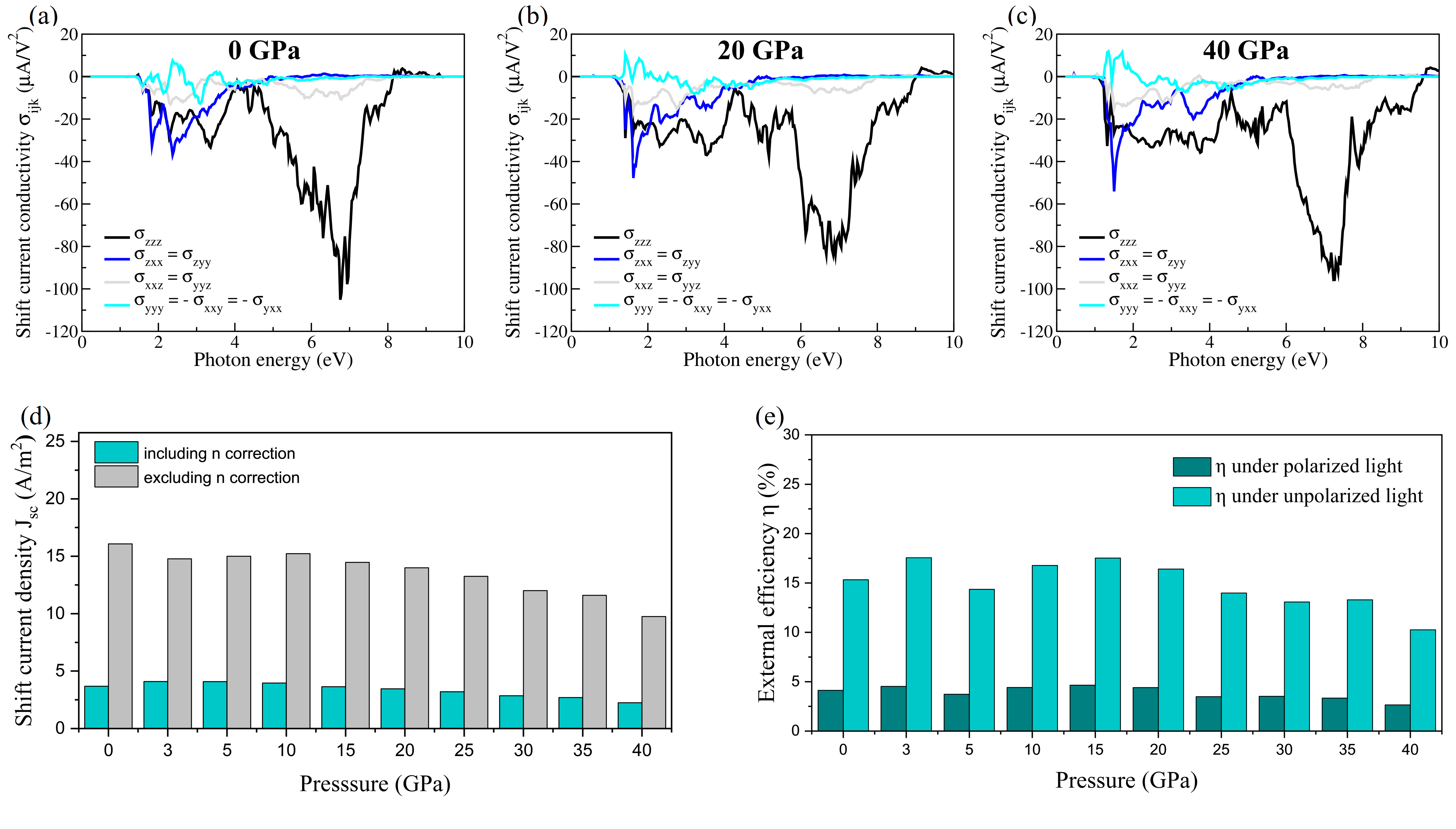}
\centering
\caption{Nonlinear optical properties of LaMoN$_{3}$ at various pressures: (a-c) Shift current conductivity ($\sigma^{\text{SC}}_{ijk}$) spectra at 0 GPa, 20 GPa, and 40 GPa, respectively. Only independent set of tensor components contributing to nonlinear optical current for \emph{R3c} symmetry are shown here. Pressure dependence of (d) shift current density (\( J_{\text{SC},z} \)) with and without refractive index (n) correction (e) shift current photovoltaic efficiency (\(\eta\)) under polarized and unpolarized illumination condition.} 
\label{Figure-4}
\end{figure*}

As such, the shift current conductivity tensor of LaMoN$_{3}$ has four independent components. We examined the shift current spectral response for all pressure-induced LaMoN$_{3}$ structures. Fig. \ref{Figure-4}(a-c) display the shift current spectra at 0 GPa, 20 GPa, and 40 GPa respectively. As shown in Eq. (9), some tensor elements have opposite signs, which indicates the potential for significant cancellations in the overall response. The dominant contributions come from the components $\sigma_{zzz}$, $\sigma_{zxx}$, and $\sigma_{zyy}$. As pressure varies, the peak magnitude of the shift current conductivity ($\sigma_{zzz}$) decreases slightly. However, in the visible energy range, the shift current conductivity, $\sigma_{zxx}$ and $\sigma_{zyy}$, increases significantly. This enhancement in the shift current response within the visible spectrum is crucial for improving the photoconversion efficiency.

The shift current density along direction $i$ in a 3D system is given by

\begin{equation}
J_{\text{SC},i} = 2 \int_0^\infty \overset{\leftrightarrow}{\sigma}^{\,\text{SC}}(\omega)\mathbf{E}(\omega)\mathbf{E}(-\omega) f(\omega) d\omega,
\end{equation}

where $\overset{\leftrightarrow}{\sigma}^{\,\text{SC}}(\omega)$ denotes the third-rank shift current conductivity tensor, $f(\omega)$ is the low-pass filter function, and the prefactor of 2 accounts for frequency permutations, following the derivation by Sipe et al.\cite{sipe2000second} The electric field $\mathbf{E}(\omega)$ represents the Fourier component of the incident field. For an ideal low-pass filter, $f(\omega) = \Theta(\omega_{co} - \omega)$, where $\Theta$ is the Heaviside step function and $\omega_{co}$ is the cutoff frequency.

In a 3D configuration, the irradiance of the incoming light relates to the time-averaged electric field as

\begin{equation}
I(\omega) = \frac{n c \epsilon_0}{2} \langle E_j^2(\omega) \rangle,
\end{equation}
where $n$ is the refractive index, $c$ is the speed of light, and $\epsilon_0$ is the vacuum permittivity. For unpolarized incident light, $\langle E_x^2(\omega) \rangle = \langle E_y^2(\omega) \rangle = \langle E_z^2(\omega) \rangle$, and mixed terms vanish. Therefore, the shift current density $J_{\text{SC},i}$ can be simplified as
\begin{equation}
J_{\text{SC},i} = 2Z_0 \int_0^\infty \sum_{j \in \{x,y,z\}} \frac{\sigma^{\text{SC}}_{ijj}(\omega) I(\omega) f(\omega)}{n(\omega)} \, d\omega,
\end{equation}

where the sum extends over all components of the orthogonal field due to the three-dimensional nature of the system and $I(\omega)$ is the spectral irradiance. $\sigma^{\text{SC}}_{ijj}(\omega)$ is shift current conductivity component that relates the incident lights electric field polarization (along j) to the shift current generated in i direction and $n(\omega)$ is the frequency-dependent refractive index.

The shift current density \( J_{\text{SC},z} \) calculated without accounting for refractive index corrections is high at pressure-dependent levels of 10-16 A/m$^{2}$ (Fig. \ref{Figure-4}d). However, incorporating the frequency-dependent refractive index into the calculation results in a substantial decrease in the current density to 3-4 A/m$^{2}$. Despite this reduction, the refractive index plays a crucial role in determining the interaction of light within the material and its absorption characteristics. Specifically, the refractive index governs the phase velocity of light and influences the spatial distribution of the electromagnetic field inside the material, which can in turn affect nonlinear optical phenomena such as the shift current. Therefore, considering the frequency-dependent refractive index is essential for accurately predicting and optimizing the bulk photovoltaic effect and overall device performance. As pressure increases, the magnitude of \( J_{\text{SC}} \) with refractive index correction steadily increases up to 10 GPa, beyond which it reduces as pressure is increased to 40 GPa.

We finally investigate how the nonlinear optical shift current influences the efficiency of the BPVE. The external efficiency of a photovoltaic device, \(\eta\), is defined as the ratio of the electrical power output from the photovoltaic device, \(P_{out}\), to the total incident energy flux, \(P_{in}\), given by
\begin{equation}
\eta = \frac{P_{out}}{P_{in}}, \quad \text{where} \quad P_{in} = \int_0^\infty I(\omega) \, d\omega,
\end{equation}
with \(I(\omega)\) representing the spectral irradiance of the incoming light. Here, \(P_{out}\) corresponds to the electrical power extracted from the photovoltaic device, while \(P_{in}\) is the total power of the incident light integrated over all frequencies. This definition provides a measure of how efficiently the device converts incident light energy into usable electrical power.

Considering that in a non-centrosymmetric system, the short circuit current is equal to the shift current \( J_{\text{SC}} \):
\begin{equation}
P_{\mathrm{out}} = F{J}_{\text{SC}} \cdot {E}_{\text{OC}}
\end{equation}
$F$ is the fill factor. The open-circuit voltage {${E}$}$_{\text{OC}}$ in the bulk photovoltaic regime is not limited by the bandgap of the absorber layer and is expressed as\cite{sauer2023shift}
\begin{equation}
E_{\mathrm{OC}} = \frac{J_{\mathrm{SC}}}{\sigma_{\mathrm{DC}}} = \frac{J_{\mathrm{SC}}}{e \left( n \mu_e + p \mu_h \right)}
\end{equation}

Considering charge neutrality condition in intrinsic semiconductors,\cite{sauer2023shift}
\begin{equation}
n = p = G \tau_r
\end{equation}
where, G is the photoexcitation generation rate given by
\begin{equation}
G = \int_0^{\infty} A(\omega) \frac{I(\omega)}{\hbar\omega} f(\omega) \, d\omega
\end{equation}

where absorptance A($\omega$) = [1 - $\exp(-\kappa(\omega) \, t)$],
$\kappa(\omega)$ is the absorption coefficient and t is the thickness of the absorber layer considered to be ~600 nm.

Thus, the external efficiency for BPVE can be expressed as, 
\begin{equation}
\eta = \frac{F \left( J_{\mathrm{SC}} \right)^2}{P_{\mathrm{in}} \, e \, G \, \tau \left( \mu_e + \mu_h \right)}
\end{equation}

Fig. \ref{Figure-4}e illustrates the external photovoltaic efficiency of LaMoN$_{3}$ calculated at all pressures under both polarized and unpolarized illumination conditions. In the case of polarized light, the efficiency exhibits a modest increase from approximately \emph{ca.} 4.1\% at ambient pressure (0 GPa) to around \emph{ca.} 4.6--5.0\% at 15 GPa. Beyond this pressure, there is a noticeable decline in efficiency, reaching approximately \emph{ca.} 2.7\% at 40 GPa. A similar trend is observed when a polarization filter is applied: the initial efficiency at 0 GPa is significantly higher at about \emph{ca.} 15.3\%, which then increases to a peak efficiency of \emph{ca.} 17.6\% at 15 GPa before decreasing to \emph{ca.} 10.3\% at 40 GPa.

A detailed examination of the efficiency expression \(\eta\) reveals that a decrease in charge carrier mobility and carrier relaxation time generally leads to an increase in the open-circuit voltage {${E}$}$_{\text{OC}}$, which in turn is expected to enhance the photovoltaic efficiency. However, despite a marked reduction in carrier mobility (Table \ref{tab:4}) with increasing pressure, the efficiency does not exhibit a corresponding significant rise. This apparent discrepancy can be explained by considering the quadratic dependence of \(\eta\) on the short-circuit current density \(J_{SC}\).
\\
The variation of \(J_{SC}\) with pressure closely mirrors the trend observed in efficiency: it increases substantially up to about 10--15 GPa, followed by a gradual decline as pressure approaches 40 GPa. The driving force behind \(J_{SC}\) is the shift current conductivity, reflecting the nonlinear optical response of the LaMoN\(_3\) system. \textcolor{black}{While spontaneous polarization contributes to this behavior, it is not the sole determining factor. Instead, the shift current is governed by a combination of structural asymmetry and electronic transition characteristics. At lower pressures, enhanced Mo–N hybridization and maintained polar distortions lead to increased optical transition probabilities and larger shift vectors, resulting in an enhanced nonlinear response. However, at higher pressures, the progressive reduction in octahedral distortion and cation off-centering weakens inversion symmetry breaking. Simultaneously, increased band dispersion, larger carrier effective masses, and modified transition matrix elements reduce the asymmetry of electronic wavefunctions. These combined effects lead to a suppression of the shift current at higher pressures despite continued bandgap reduction. Thus, the observed peak in shift current around 10–15 GPa arises from an optimal balance between structural asymmetry and electronic transition strength.}
\\

Overall, the linear optical response (governed by the dielectric function and absorption coefficient) determines the photon absorption and carrier generation efficiency, which directly contributes to the SLME - a key indicator of conventional photovoltaic (PV) performance. Under pressure, LaMoN$_{3}$ exhibits an increased absorption coefficient and reduced bandgap, leading to higher SLME values, thereby improving the linear PV conversion efficiency.
In contrast, the nonlinear optical response, particularly the BPV or shift-current effect, generates photocurrents in non-centrosymmetric materials without requiring a p-n junction. This effect originates from the asymmetric excitation and shift of photoexcited carriers. Our results reveal a higher shift current density up to 15 GPa, marking an optimal regime for nonlinear photovoltaic response.
Therefore, by harvesting both linear (SLME-driven) and nonlinear (shift-current-driven) mechanisms, LaMoN$_{3}$ can utilize the full potential of light-matter interaction: the 40 GPa phase favors efficient linear absorption in thicker films, while the 15 GPa phase maximizes nonlinear shift current in nanoscale absorbers. Together, these regimes offer a synergistic pathway to enhance overall optoelectronic performance. This dual mechanism is a core novelty of the work.

\section{Conclusion}

In summary, we conducted a systematic investigation of the pressure--dependent phase stability, electronic structure, optical response, excitonic and polaronic effects, as well as shift-current properties of non--centrosymmetric ferroelectric LaMoN$_{3}$ using {\it ab-initio} density functional calculations. LaMoN$_{3}$ remains dynamically stable and retains its single--phase structure up to 40 GPa, exhibiting an indirect band gap that decreases from 2.17 eV at 0 GPa to 1.45 eV at 40 GPa within the G$_{0}$W$_{0}$@PBE framework. Pressure also enhances the SLME from 8.66\% to 26.38\% and lowers the exciton binding energy from 184 meV to 37 meV, both trends favorable for optoelectronic applications. In contrast, pressure strengthens carrier--phonon interactions, thereby increasing the polar character while reducing carrier mobility. \textcolor{black}{The intrinsic polar nature of LaMoN$_{3}$ drives spontaneous polarization, which underlies its shift--current generation; the shift--current response increases up to 15 GPa but diminishes at higher pressures. However, the reduction in shift current magnitude cannot be attributed solely to decreased polarization; rather, it reflects changes in electronic structure due to underlying pressure. Although shift current and polarization share a common symmetry origin, the former is governed by the shift vector which is linked to the difference in Berry-phase polarization between valence and conduction bands. Although moderate compression enhances orbital overlap and optical transitions, higher pressures lead to band broadening, increased carrier effective mass and a reduced shift vector that ultimately diminishes the NLO response.}

Building on our comprehensive investigation, we further analyzed the bulk photovoltaic efficiency of LaMoN$_{3}$ under varying pressure conditions and illumination polarizations to capture nonlinear optical contributions. The pressure dependence of efficiency closely follows the short-circuit current density \(J_{SC}\) behavior, peaking near 15 GPa due to the enhanced nonlinear shift-current response, and subsequently declining at higher pressures as nonlinear optical currents is reduced. At 40 GPa, the optimal power conversion efficiency is achieved for absorber thicknesses on the order of 100 $\mu$m, whereas at 15 GPa, maximum device performance occurs in ultrathin layers (t = $\sim$ 600 nm), optimized for nonlinear current generation. Importantly, the 40 GPa phase exhibits a narrowed band gap of ~1.5 eV, favoring linear optical absorption in the lower photon-energy region, while the 15 GPa phase retains a wider band gap of ~2.0 eV, enabling efficient capture of higher-energy photons through nonlinear processes. This complementary spectral behavior highlights the potential of LaMoN$_{3}$ to simultaneously exploit linear and nonlinear photovoltaic currents in multi-junction device architectures.

Leveraging these pressure-tuned phases, multi--junction architectures can be designed to exploit distinct spectral regions and optical mechanisms, thereby maximizing overall photovoltaic performance. This strategy resonates with prior studies on complex oxide photovoltaics, such as Bi$_2$FeCrO$_6$, where compositional and structural tuning has enabled multi--bandgap solar cells with enhanced bulk photovoltaic effects\cite{Alexe2015}. Similarly, state--of--the--art III--V multijunction solar cells have long demonstrated the effectiveness of stacking materials with complementary band gaps to surpass single-junction efficiency limits\cite{Raisa2024,Heydarian2024,Baiju2022}. Anisotropic strain in LaMoN$_{3}$ can be achieved practically at ambient conditions by combining non-pressure methods that independently modify each crystallographic direction, for example, epitaxial growth on lattice-mismatched substrates for in-plane strain and nanostructuring or thermal quenching treatments to tune out-of-plane strain.\cite{sun2018three,cao2021vertical,badding1995high} Our results establish a foundation for the rational design of nitride perovskite--based multi--junction solar cells, where pressure--induced bandgap engineering enables the synergistic capture of both linear and nonlinear optical currents, thereby advancing the development of high--efficiency, multifunctional photovoltaic devices.

\section*{Supporting Information}
Calculation details of phonon dispersion curves, mathematical formulation of the acoustic sum rule, chemical phase diagrams, effect of SOC on the band structure, G$_0$W$_0$@PBE band structures, single phonon angular frequency ($\omega_{LO}$) and minimization of the free polaron energy, spectroscopic limited maximum efficiency (SLME), and non-linear optical properties of LaMoN$_3$ under different applied pressures.

\section*{Author Contributions}
S.A and S.S.P contributed equally to this work.

\section*{Data availability}
The authors confirm that the data supporting this article have been included as a part of the Supporting Information. Derived data supporting the findings of this study are available from the authors upon reasonable request.

\section*{Conflicts of interest}
There are no conflicts to declare.

\section*{Acknowledgements}
A.A, S.A, and S.S.P acknowledges computing facility (spacetime2, Paramrudra) provided by IIT Bombay to support this research. A.N.S, J.J, and S.S.P acknowledges the computational support by MASSIVE HPC facility (www.massive.org.au) and the Monash eResearch Centre and eSolutions-Research Support Services through the use of the MonARCH HPC Cluster. A part of this research was undertaken with the assistance of resources from the National Computational Infrastructure (NCI Australia), an NCRIS enabled capability supported by the Australian Government. J.J. acknowledges funding through the Australian Centre for Advanced Photovoltaics funded by ARENA. 
	




\bibliography{ACS}
\end{document}